\documentclass[sigchi, urlbreakonhyphens=false, screen]{acmart}

\copyrightyear{2023}
\acmYear{2023}
\setcopyright{acmlicensed}
\acmConference[CHI '23]{Proceedings of the 2023 CHI Conference on Human Factors in Computing Systems}{April 23--28, 2023}{Hamburg, Germany}
\acmBooktitle{Proceedings of the 2023 CHI Conference on Human Factors in Computing Systems (CHI '23), April 23--28, 2023, Hamburg, Germany}
\acmPrice{15.00}
\acmDOI{10.1145/3544548.3581060}
\acmISBN{978-1-4503-9421-5/23/04}

\usepackage{booktabs}
\usepackage{graphicx}

\usepackage[english=american,threshold=40,thresholdtype=words]{csquotes}

\newcommand{\surveyTitle}{\textbf{End-User Participant Perspectives}}
\newcommand{\interviewTitle}{\textbf{Developer Participant Perspectives}}

\begin{document}

\title[Developer \& End-User Perspectives on App Permissions \& Their Privacy Ramifications]{Stuck in the Permissions With You: Developer \& End-User Perspectives on App Permissions \& Their Privacy Ramifications}

\author{Mohammad Tahaei}
\orcid{0000-0001-9666-2663}
\email{mohammad.tahaei@nokia-bell-labs.com}
\authornote{Contributions were made while at the University of Bristol.}
\affiliation{
    \institution{Nokia Bell Labs}
    \country{United Kingdom}
}

\author{Ruba Abu-Salma}
\orcid{0000-0002-5316-9956}
\email{ruba.abu-salma@kcl.ac.uk}
\affiliation{
    \institution{King's College London}
    \country{United Kingdom}
}

\author{Awais Rashid}
\orcid{0000-0002-0109-1341}
\email{awais.rashid@bristol.ac.uk}
\affiliation{
    \institution{University of Bristol}
    \country{United Kingdom}
}

\begin{abstract}
While the literature on permissions from the end-user perspective is rich, there is a lack of empirical research on why developers request permissions, their conceptualization of permissions, and how their perspectives compare with end-users' perspectives. Our study aims to address these gaps using a mixed-methods approach.

Through interviews with 19 app developers and a survey of 309 Android and iOS end-users, we found that both groups shared similar concerns about unnecessary permissions breaking trust, damaging the app's reputation, and potentially allowing access to sensitive data. We also found that developer participants sometimes requested multiple permissions due to confusion about the scope of certain permissions or third-party library requirements. Additionally, most end-user participants believed they were responsible for granting a permission request, and it was their choice to do so, a belief shared by many developer participants. Our findings have implications for improving the permission ecosystem for both developers and end-users.
\end{abstract}



\maketitle

\newpage
\section{Introduction}
\label{sec:intro}
Permissions are the primary mechanism for protecting data and resources on smartphones, including end-users' location, contacts, and photos. Developers use permissions to ask for data and resources needed for particular app features to function, and end-users, on the other hand, decide whether they want to grant or deny such requests.\footnote{Throughout the paper, we refer to \textit{developers} as those who develop smartphone apps and \textit{end-users} as those who use smartphone apps.} Developers play a critical role in choosing which app permissions to include or exclude and, consequently, in the app ecosystem. However, despite the rich literature on permissions, there have been no empirical studies \textit{with} developers about their views, understanding, and practices with regard to permissions and how these compare to end-users' attitudes toward permissions.

End-users primarily grant or deny permission requests based on how permissions are related to the app's functionality and end-users' expectations~\cite{wijesekera2015android, shen2021can, elbitar2021explanation}. Some end-users also consider the privacy ramifications of permissions and make contextualized decisions~\cite{wijesekera2015android}. However, developers' decision-making processes with regard to permissions and how developers' perspectives on permissions compare with end-users' perspectives have not been studied. Besides the lack of research on developers' understanding of permissions, the recurring pattern of unused and excessive permissions requested by apps over the past ten years~\cite{felt2011demystified, mallojula2021you, wang2022aper} as well as end-users' privacy concerns about permissions (e.g., data leaks to third parties~\cite{wang2022aper, reardon2019ways,hallinan2021data}) emphasize the need for understanding how developers decide on which permissions to include or exclude in their apps and what challenges they face when integrating permissions.

In this study, we shed light on the decision-making processes of (and the challenges faced by) developers with regard to permissions. We also augment our findings with a follow-up study with end-users to give in-depth insights into the permission ecosystem from the perspectives of two primary stakeholders of the app ecosystem, developers and end-users. Our research questions (RQs) are: 

\begin{enumerate}
    \item[\textbf{RQ1:}] How do developers decide on what permissions to request?
    \item[\textbf{RQ2:}] What is developers' understanding of permissions and permissions' privacy ramifications?
    \item[\textbf{RQ3:}] What are end-users' attitudes toward permissions, and how do these compare with developers' views of permissions?
\end{enumerate}

We conducted a mixed-methods study with developers and end-users to answer our RQs. We interviewed 19 developers about their practices with regard to permissions, decision-making processes, and challenges when integrating permissions (RQ1 \& RQ2). Based on these findings, we then designed and conducted a survey with 309 Android and iOS end-users to explore their views on permissions and compare them with developer participant views (RQ3).

We found that developer participants often viewed permissions as an access control mechanism, a perspective shared by some of our end-user participants. Some end-user participants viewed permissions as a legal guard against the misuse of their data, a view not shared by developer participants who rarely brought up legal ramifications of permissions. We also found that app features and functionality were the primary reasons for developer participants to include permissions. However, they rarely removed permissions. Further, developer participants were aware of the poor end-user experience of requesting too many permissions, and they highly regarded the trust and reputation that came with respecting end-users. Our results show that some developer participants needed clarification on the scope of permissions (mainly when a resource could be accessed with different permissions, e.g., fine location vs. coarse location). Due to this confusion, they sometimes added multiple permissions to avoid crashes or unexpected consequences. Additionally, our developer participants included third-party libraries not only for providing functionality but also for helping them manage permissions, which sometimes resulted in apps requesting permissions that developer participants did not expect.

Like developer participants, our end-user participants mostly granted permissions because of features, and some granted permissions because they trusted the app. 31.4\% of our end-user participants did not see any harm in permissions, and those who saw harm in granting permissions were worried about unintended use of their data, such as selling their data to third parties. 57.9\% of our end-user participants had never removed already granted permissions because they believed they had no reason to do so, did not know how to do so, or had not thought about doing so previously.

\section{Related Work}
\label{sec:related-work}

\subsection{App Permissions}
Based on analyses of Android app reviews~\cite{nema2022analyzing, harkous2022hark, nguyen2019short}, one of the main privacy concerns of end-users relates to app permissions and what access apps have to data and resources on their smartphones. However, when a permission request is contextualized, matches end-users' expectations, and is perceived by end-users as necessary for an app to function, end-users are likely to grant the permission request~\cite{malkin2022runtime, shen2021can, micinski2017user, elbitar2021explanation, wijesekera2015android}, which is rooted in the theory of Contextual Integrity~\cite{wijesekera2015android}. For example, an end-user may understand why service providers need to gather certain data about end-users as long as providers protect and do not misuse end-user data~\cite{malkin2022runtime}. Therefore, knowing why a developer includes or excludes a permission request can shed light on a primary privacy concern of end-users, permission requests.

Unused and excessive permissions have been a recurring pattern in apps over the past ten years~\cite{felt2011demystified, felt2012how, mallojula2021you}. Studying developer artifacts suggests that developers may leave unused or unnecessary permissions in their code due to updating the list of permissions requested in different app development stages (e.g., a developer may include a permission request during the testing stage and forget to remove it later during the production stage)~\cite{scoccia2019permissions, felt2011demystified}, copy-pasting code from the Internet, lack of access to comprehensive documentation, confusion about permissions' naming, and not fully understanding the scope and use cases of permissions~\cite{felt2011demystified}. On the positive side, developers can be nudged toward reducing the number of permissions they include by knowing that apps with similar functionalities ask for fewer permissions. Such a nudge in the form of a warning message that pops up when developers submit their apps to the Google Play Store has improved developers' decisions and reduced permissions~\cite{peddinti2019reducing, google2020helping}. However, using unnecessary permissions or forgetting to remove unused permissions remains a problem~\cite{mallojula2021you, calciati2020automatically, scoccia2019permissions}. These permissions can stay in an app, shifting the responsibility to end-users to review them at some point later in time (if granted), which is not a typical behavior among end-users~\cite{shen2021can}. Apps with unnecessary permissions can cause end-user frustration and have privacy ramifications for them~\cite{felt2011demystified, felt2012user}.

Despite the importance of permission systems in the app privacy ecosystem and the plethora of studies with end-users, the \emph{only} study exploring issues faced by developers when working with permissions had explicitly focused on iOS permission descriptions in 2014, when these descriptions were in their early stages, using a survey~\cite{tan2014effect}. Developers had a mixed understanding of why descriptions were needed and were unsure how to write them, and some were also unaware of such descriptions. The permission ecosystem has changed significantly since the study was conducted; for example, Android has introduced run-time permissions instead of install-time permissions (ask at install time vs. ask when needed)~\cite{google2022permissions}.

While there is a rich literature on permissions, there have been no empirical studies \emph{with} developers about permissions. Our study bridges the gap between how developers, on the one hand, and how end-users, on the other hand, understand permissions. We provide empirical insights into developers' understanding and decision-making processes with regard to app permissions using interviews. Drawing on our interview findings, we design and conduct a survey of end-users to capture their perspectives on permissions and compare them with developer perspectives.

\subsection{Empirical Privacy Studies With Developers}
A strand of empirical software engineering research has studied the support developers need to build privacy-friendly apps and perform privacy-related tasks~\cite{tahaei2019survey, tahaei2022embedding}, such as finding privacy issues in code~\cite{li2018coconut}, deciding on personalizing ads~\cite{tahaei2021deciding}, and building apps that are compliant with privacy laws~\cite{tahaei2022charting, alomar2022developers}. In the case of making child-directed apps compliant with privacy laws like the California Consumer Privacy Act (CCPA)~\cite{coppa2022}, developers often try to satisfy app store requirements instead of laws, and they often rely on app stores and operating systems to detect privacy-related issues~\cite{alomar2022developers, tahaei2022embedding}.

Third-party libraries are often confusing for developers because of the libraries' unclear data collection practices and complicated configurations~\cite{alomar2022developers}. These libraries could be unclear about their purposes and permission requests, causing developers to send unnecessary sensitive data about end-users to third parties without developers' knowledge~\cite{reardon2019ways,razaghpanah2018apps}. Studies of privacy posts on developer forums such as Stack Overflow, Reddit, and iPhoneDevSDK have shown that managing permissions is a complex task for developers~\cite{li2021how, tahaei2022advice, tahaei2020so, shilton2019linking}, due in part to developers not understanding the scope of specific permissions or, more broadly, the app ecosystem~\cite{li2021how}. On the other hand, experienced developers recognize the direct relationship between asking for permissions and end-users trusting an app, recommending fellow developers to ask for permissions only when necessary, with clear permission descriptions~\cite{shilton2019linking}.

In this work, we contribute to this body of research by providing first-hand qualitative insights into developers' understanding of and decision-making processes with regard to permissions using in-depth interviews.

\section{Method}
To answer our RQs (\S\ref{sec:intro}), we first conducted 19 interviews with developers to understand their thought processes with regard to permissions (RQ1 \& RQ2). Our interview findings informed the design of a survey study we conducted with 309 smartphone end-users. The survey aimed to explore end-users' perspectives on permissions and compare them with those of developers (RQ3). The Research Ethics Committee at the University of Bristol (Faculty of Engineering) reviewed and approved our study. 

\subsection{Interview Study With Developers} 
\label{sec:interview-devs}

\subsubsection{Developer Participant Recruitment}
We advertised our interview study on LinkedIn~\cite{linkedin2022}, Twitter~\cite{twitter2022}, and two freelancing websites, namely Freelancer~\cite{freelancer2022} and Upwork~\cite{upwork2022}. We also used Prolific~\cite{prolific2022}, a crowdsourcing platform, to find additional developer participants. On Prolific, we invited interested participants who stated they had programming skills and worked in the computer and information technology industry. All these approaches and platforms were used in the literature to recruit developers~\cite{tahaei2021sat,votipka2020scale,tahaei2022charting,gutfleisch2022how}. We recruited developer participants between April 2022 and May 2022.

\paragraph{Screening Survey}
To ensure that all developer participants had a background in programming, app development, and permissions, we asked all interested candidates to fill out a short survey describing their role in their last software development job, years of experience in software development, and app development. We also included six programming questions, suggested by \citeauthor{danilova2021do}~\cite{danilova2021do}, to screen out participants who did not have basic programming knowledge. The first four programming questions were about the definition of a compiler function, the possible value of a Boolean variable, the website that programmers frequently visited, and the definition of a recursive function. Two additional questions assessed participants' understanding of a short pseudocode snippet. The screening survey can be found in Appendix~\ref{app:screening-survey}.

It took interested candidates five minutes on average (SD=2) to complete the screening survey. After screening out those who did not pass the programming questions, we ensured that those who passed had at least two years of software development experience to set a baseline for participants' experience with permissions. In total, 67 developer participants met our criteria to be invited to our interview study; 19 participated. The interview transcripts are unavailable online due to privacy and ethical considerations; however, they can be accessed upon request for research purposes.

\subsubsection{Developer Participant Demographics}
We interviewed 19 developer participants via audio calls; eight were from Upwork, six from Freelancer, three from LinkedIn, and two from Prolific. On average, they were thirty years old (SD=6), and they had eight years of experience in software development (SD=5), four years of experience in Android development (SD=3), and three years of experience in iOS development (SD=3). Fourteen self-identified as male and three as female, and two preferred not to describe their gender. Ten were located in Asia, three in Europe, two in Africa, two in North America, and one in South America. They all had experience working with permissions. Table~\ref{tab:demographics} in Appendix~\ref{app:demographics-developer} summarizes developer participants' demographics.

\paragraph{App Development Experience} 
Participants developed different apps, including healthcare, utility, e-commerce, educational, social media, gaming, and finance apps. Four participants developed apps specifically targeted at adults and four at kids; the rest did not have age limits for their apps.

When developing apps, participants reported that they collected different types of data mainly depending on what their apps required to function correctly, including end-users' phone numbers, email addresses, countries of residence, locations at specific times, photos, videos, contacts, messages, device IP addresses and tokens (for fingerprinting), purchase history, shopping preferences, and data for debugging, analytics, and measuring performance. Developer participants also used different libraries when developing apps. The most common ones were Firebase, Google Analytics, Flurry, Retrofit, and other libraries for image processing, QR code scanning, data visualization, and crash report generation.

\subsubsection{Study Procedure}
After obtaining our developer participants' consent and briefly explaining the study (we sent participants the consent form and additional information about the study ahead of the interviews), we asked participants about their job background and experience in app development. We then explored their understanding of permissions, decision-making processes, whether they updated (by adding or removing) permissions post-development, and their confusing and challenging experiences with permissions.

Based on their platform expertise, we showed participants the 15 most commonly used permissions in Android, iOS, or both, taken from \citeauthor{kollnig2022are}~\cite{kollnig2022are} and asked them to describe what they thought those permissions did. We also discussed what data they considered sensitive, what privacy within the context of app permissions meant to them, whether they were aware of any privacy laws, and, if so, how they complied with them when choosing permissions. Through these discussions, we explored participants' understanding of permissions' harms and privacy ramifications. Interviews ended with questions about participants' ideas or recommendations for improving permissions, as well as any other missing thoughts participants might have.

The interviews took, on average, 44 minutes (SD=10). We reached data saturation; the interviewer frequently compared notes on the topics arising and discussed with another author whether or not to continue interviews after every few. Each developer participant received £30 for their time.\footnote{Although our developer participants did not reside in the U.K. (Table~\ref{tab:demographics}), as per the guidelines and recommendations provided by our institution's Research Ethics Committee, we needed to comply with domestic labor laws in the U.K. Hence, we paid our developer participants about three times the U.K. National Minimum wage, which was £9.5 at the time of recruitment; developer participants received £30).} All interviews were conducted in English. The interview guide can be found in Appendix~\ref{app:interview-guide}.

\subsubsection{Data Analysis}
All interviews were audio recorded and transcribed using professional privacy-compliant transcription services. We then imported the interview transcripts to NVivo for analysis~\cite{nvivo2022}. Two authors inductively coded all 19 transcripts using thematic analysis~\cite{miles1994qualitative,saldana2015coding}. To develop an initial codebook, they first independently open-coded two transcripts (the same ones) and developed their own codebooks. In multiple sessions, they discussed their codebooks, merged them into one, and resolved disagreements. They also sought additional feedback and input from a third author on their codebook, notes, and preliminary findings. The codebook-building procedure was iteratively followed until the codebook was stable, signaling that code saturation was achieved; new codes stopped emerging. Code saturation occurred after 11 iterations---each iteration involved both authors independently coding one transcript (the same transcript but different from transcripts used in other iterations), meeting to resolve disagreements, and refining the codebook. Both authors coded two additional transcripts after the $11^{\text{th}}$ iteration using the final codebook to ensure the codebook was stable and no changes were made. Codes were not mutually exclusive, and a quote might appear in multiple codes.

The two authors then independently coded all 19 transcripts using the final stable codebook and measured inter-rater reliability using NVivo. The average Cohen's kappa coefficient across all codes was .71, which is considered a substantial agreement~\cite{landis1977interrater}. Both authors resolved the remaining disagreements through discussions. The qualitative findings in \S\ref{sec:developer-findings}, labeled with \textbf{\textit{Developer Participant Perspectives}}, are based on the final codebook and resolved disagreements. Due to the qualitative nature of our interview study, we do not report frequencies of occurrences in \S\ref{sec:developer-findings}. We instead use qualifiers (e.g., few, some, several, many, all) to avoid overgeneralization~\cite{maxwell2010using}. However, Table~\ref{tab:codebook} in Appendix~\ref{app:codebook} includes the codebook with frequencies for interested readers.

\subsection{Survey Study With End-Users} \label{sec:survey-users}

\subsubsection{End-User Participant Recruitment}
We recruited 309\footnote{Prolific's minimum size of a representative sample of the U.K. is 300~\cite{prolific2022sample}.} end-user participants using Prolific~\cite{prolific2022}, a typical crowdsourcing platform for recruiting participants for empirical privacy studies~\cite{distler2021systematic}. We decided to recruit participants residing in the U.K. because (1) privacy is a cultural and contextual topic, and people's interpretation of what privacy means can vary based on where they are located~\cite{sambasivan2018privacy, yang2016examining, busse2020cash,sawya2017self};\footnote{We did not consider this for our interview study with developers because our sample was small, and our developer participants developed apps for end-users across the globe and not necessarily for end-users in the developers' home countries.} (2) Prolific offers representative samples only for two countries, the U.S. and the U.K.~\cite {prolific2022sample}; (3) we had limited resources and budget; and (4) a simple translation from English into other languages could have caused different interpretations of our questions. To avoid this, we would have needed to perform a validated translation, instead of a simple translation, for consistency (as stated in~\cite{sawya2017self}), requiring additional resources. Thus, we ran our survey in English.

We used Prolific's prescreening feature to target participants residing in the U.K. (for the reasons stated above) and using Android or iOS because these two had over 99\% of the market share of all smartphone operating systems~\cite{statista2022mobile}. We also used a gender-balanced sample provided by Prolific because it could generate an almost similar sample to a representative sample with a lower cost~\cite{tang2022replication}.

We ran the survey in August 2022 and paid participants £1.84, slightly above the minimum hourly wage in the U.K. (£9.50, all survey participants were located in the U.K.). It took participants, on average, 11 minutes (SD=5) to complete the survey. 
Our anonymized survey dataset can be accessed upon request for research purposes only at \url{https://doi.org/10.5523/bris.2qv93u6isblq42ixbofaahnfwm}.

\subsubsection{End-User Participant Demographics}
Our end-user participants were all located in the U.K. On average, they were 37 years old (SD=13). 50.2\% self-identified as male and 49.8\% as female. 38.3\% were employed full-time, 14.2\% were employed part-time, 9.5\% were not in paid work, 7.4\% were unemployed (and job seeking), 1.2\% were due to start a job within the next month, and 29.4\% did not have an updated employment status on Prolific. On average, our end-user participants spent 5 hours (SD=3) on their smartphones daily and had a smartphone for 11 years (SD=4). 51.1\% were Android end-users, and 48.9\% were iOS end-users.

\subsubsection{Study Procedure}
The survey consisted of open-ended and closed-ended questions inspired by our interview study. The open-ended questions explored end-user participants' understanding of permissions and their perceptions of the harms associated with permissions, as discussed with developer participants. The close-ended questions assessed whether the assumptions our developer participants made about end-users were accurate (e.g., end-users did not care about or were not aware of permissions, see \S\ref{sec:assumptions}), end-users' opinions of the privacy ramifications of commonly used permissions in Android and iOS (e.g., photos, locations, and camera)~\cite{kollnig2022are}, how end-users interacted with permissions, and how they felt about permissions. Where possible and relevant, we also adapted questions from prior surveys~\cite{tang2022replication, abrokwa2021comparing, harborth2021evaluating, kariryaa2021understanding, ismail2017permit, degirmenci2020mobile, bonne2017exploring}. Our survey instrument can be found in Appendix~\ref{app:end-users-survey}.

\subsubsection{Data Analysis}
We did a descriptive analysis of responses to all closed-ended (multiple-choice \& Likert) questions. For the open-ended questions, two authors collaboratively analyzed the qualitative responses using Miro~\cite{miro2022} boards. We copied all responses into Miro using sticky notes and constructed groups based on the similarities across responses (i.e., building affinity diagrams~\cite{braun2006thematic, lazar2017interviews}). Some notes ended up in multiple groups, and groups were not mutually exclusive. The subsections labeled with \textbf{\textit{End-User Participant Perspectives}} in \S\ref{sec:developer-findings} are based on these analyses.

\subsection{Limitations}
\label{sec:limitations}
Our samples do not necessarily represent all developers and end-users. However, we recruited our developer participants using different channels and platforms to reduce sampling bias. Nonetheless, most participants in our sample self-identified as male (consistent with the gender-biased software development profession; over 90\% of software developers are male~\cite{statista2020gender}) and were from Asia. Hence, our findings are not generalizable to all developers. However, they provide novel insights into developers' understanding and practices with regard to permissions and their privacy ramifications, which is the main objective of our qualitative research. Similarly, our survey targeted a Western country: the U.K. Future research may use our public data to compare the results obtained from end-user participants residing in other countries with our findings. Furthermore, our study used a mixed-methods design: an interview-based study and a survey; each was appropriate for the targeted participant population (developers and end-users, respectively). Therefore, our results should be interpreted with these limitations in mind.

We did not run a survey with developers. First, unlike an interview-based study, a survey would have prevented us from generating in-depth insights into developers' understanding and practices. Second, due to privacy and ethical considerations, we could not recruit participants by harvesting their email addresses from software development platforms and forums (e.g., GitHub and Google Play) or sending unsolicited emails to developers to take part in our study (for sampling issues, see~\cite{UtzEtAl2022, tahaei2022recruiting, tahaei2022lessons}). For future work, researchers with access to a large pool of developers could run a survey with developers to collect more data and compare their data with our publicly available survey data. Besides, we did not collect data about our developer participants' apps (e.g., magnitude of the app's end-user base and popularity, as well as the category of the app) or development mode (e.g., employed by an agency to develop apps customized for clients, working in a company on a single app, or working as a freelancer), which could have helped contextualize our findings. We suggest that future researchers gather this data as part of their screening or demographics survey.

Although we did not aim in this study to explore developers' privacy compliance processes, we asked our developer participants whether they had made changes to their permission requests in order to make their apps compliant with privacy laws. They all noted that they did not make any changes (\S\ref{sec:os-appstore}). A recent study has found that child-directed app developers either assume that their apps are compliant as long as they have not been rejected by app stores or outsource most compliance decisions to auditing services~\cite{alomar2022developers}. Future research may want to investigate to what extent end-users expect developers to be aware of privacy laws and cater to legal necessities, and how different privacy laws may impact the permission requests, especially since some of our end-user participants (19\%) perceived requesting permissions as a privacy compliance mechanism (see \S\ref{sec:why-perms} for details).

\section{Findings}
\label{sec:developer-findings}

\begin{figure*}
  \centering
  \includegraphics[width=.9\linewidth]{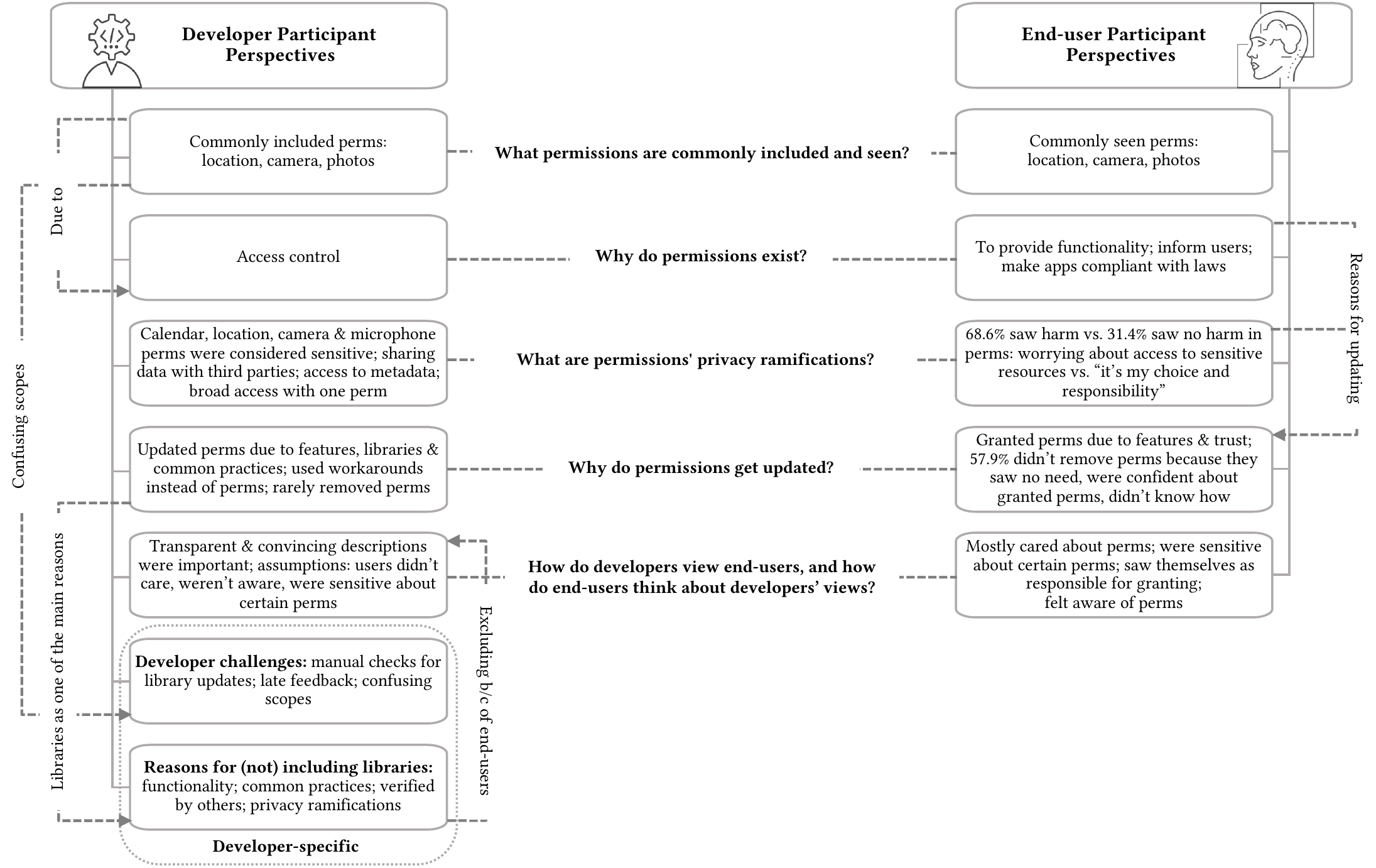}
  \caption{An overview of our findings from the interview with developer participants and the survey with end-user participants.}
  \Description{An overview of our findings from the interview study with developer participants and the survey study with end-user participants, showing two stakeholders, developers and end-users, on each side of the figure. Findings are laid out in boxes, and the themes from the interview analysis are centered in the middle.}
  \label{fig:overview}
\end{figure*}

Figure~\ref{fig:overview} shows a summary of our findings. In each subsection, we first outline our findings from the interview study with developer participants (RQ1 \& RQ2, \textbf{\textit{Developer Participant Perspectives}}). We then integrate the findings from the survey with end-user participants in these subsections (RQ3, \textbf{\textit{End-User Participant Perspectives}}), where appropriate. \S\ref{sec:developer-challenges} \& \S\ref{sec:reasons-libraries} cover topics specific to developer participants.

\subsection{Commonly-Used Permissions}
\label{sec:use-of-permissions}
\subsubsection*{\interviewTitle} Our developer participants mostly requested access to end-users' location, camera, photo gallery, Internet, Bluetooth, data storage, contacts, messages, and microphone. They understood what most Android and iOS permissions did (e.g., Internet or access to storage permission). However, some \textbf{could not differentiate} between permissions with similar scopes. Some Android developers were \textbf{confused} about the network and Wi-Fi permissions, relating them to learning about the current connection. Similarly, many iOS developers were unsure whether they could access all or specific photos using PhotoLibrary and PhotoLibraryAdd (see \S\ref{sec:confusing-scopes} for the privacy ramifications due to this confusion). Similar confusions occurred when permissions provided access to foreground or background services (e.g., location).

\subsubsection*{\surveyTitle}
The location permission request, with 98.1\% of end-user participants seeing it in the apps they had frequently used in the past year, was the most requested permission from our end-users' perspective (Appendix~\ref{app:plots-survey}, Figure~\ref{fig:fivePerms}). While the location permission is not the most commonly used in Android and iOS (with evidence from an analysis of apps~\cite{kollnig2022are}), end-user participants might remember the location permission request the most because they were sensitive about it or because their commonly used apps often required this permission. 

The camera (82.5\%), photos (81.6\%), microphone (64.1\%), and contacts (54.7\%) permissions were next in line for commonly seen permission requests by our end-user participants. Other permissions, such as the Internet and network state that did not require a run-time permission request from the end-user, were mentioned less often (19.7\% and 19.1\%, respectively).

\subsection{Why Do Permissions Exist? What Do They Do?}
\label{sec:why-perms}

\subsubsection*{\interviewTitle}
Most developer participants used technical terms revolving around the concept of \textbf{access control}, which is not far from why apps request permissions (e.g., iOS describes permissions as ``control[ing] access to information in apps on iPhone.''~\cite{apple2022permissions}). Our developer participants used the following concepts to explain permissions: a \textbf{switch} used to enable access to specific restricted resources; a \textbf{bridge} between developers and end-users; a \textbf{medium} to protect end-users and their devices from malicious activities; and an end-user giving \textbf{consent} to accessing a resource they owned.

\subsubsection*{\surveyTitle} Most end-user participants correctly identified what permissions did: granting access to specific resources or data on their smartphone (87.7\%)---similar to the access control concept from the developer study. Regarding apps that end-user participants had used in the past year, at least 70\% of end-user participants agreed that they were \textbf{aware} of permissions used by these apps, the number of permission requests was \textbf{in line with their expectations}, and they \textbf{understood} why certain permissions were requested (Figure~\ref{fig:fiveAppsLikertsLikerts}). Unlike our observation, the findings of a recent comprehension survey study with end-users have shown that most end-user participants could not understand the scope of several permission requests~\cite {shen2021can}. We speculate that this discrepancy may be attributed to the studies' methods: our data was self-reported, whereas \citeauthor{shen2021can}'s data was based on a knowledge survey. Hence, although most of our end-user participants reported being aware of permissions, this does not mean they fully understood the permissions' scope. Further, end-user participants used lay descriptions, compared to developer participants, to explain why apps requested permissions, as detailed below:

\begin{figure*}
  \centering
  \includegraphics[width=.9\linewidth]{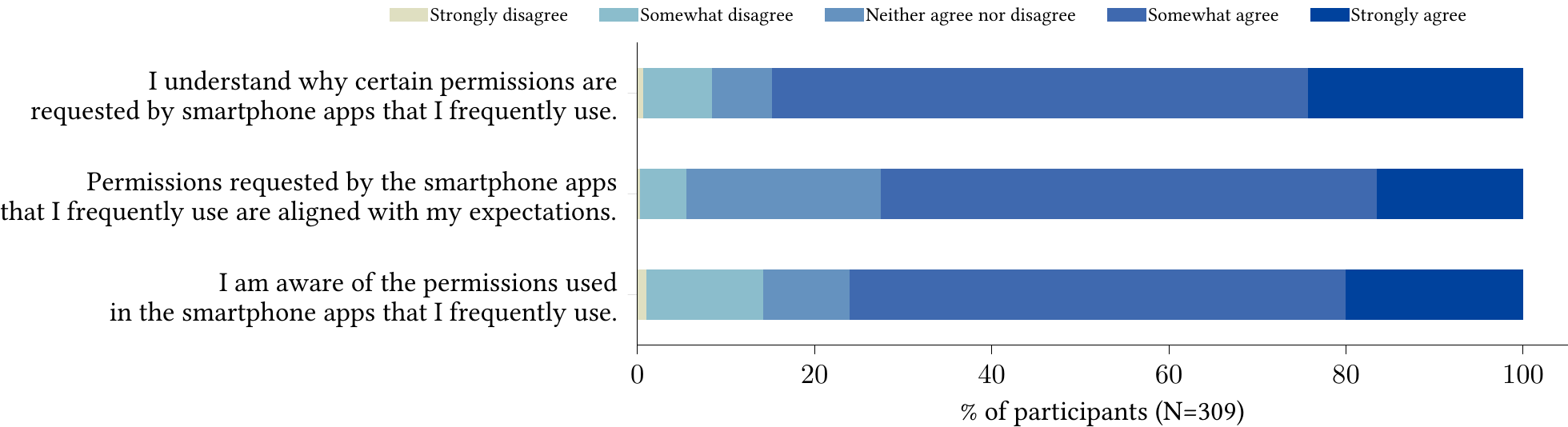}
  \caption{End-user participants' answers to ``Thinking about the five smartphone apps that you have frequently used in the past year, to what extent do you agree or disagree with the following statements?''}
  \Description{End-user participants' answers to ``Thinking about the five smartphone apps that you have frequently used in the past year, to what extent do you agree or disagree with the following statements?'' Each bar shows end-users' answers to a Likert question.}
  \label{fig:fiveAppsLikertsLikerts}
\end{figure*}

\begin{itemize}
    \item To help with app's \textbf{functionality} (35\%): in line with our developer participants' reasons for granting permissions and prior research with end-users~\cite{wijesekera2015android}, our end-user participants' predominant understanding of permissions was to provide some functionality, such as storing files, taking photos, and finding a location;
    
    \item To \textbf{inform} end-users (21\%): this theme covered topics around informing and notifying end-user participants of the rationale behind the permission request and giving them the option to accept or deny the request. Participants believed that the control they received from the permission request could limit access to their data or stop random apps from being installed on their smartphones;
    
    \item To make apps \textbf{compliant} with laws (19\%): while a few developer participants discussed compliance with privacy and data protection laws, some end-user participants viewed permissions as a way for apps to comply with laws. They thought permissions helped developers feel protected against failing to comply with laws when requesting end-user data. They viewed permission requests as part of legal practices that developers must follow---using a smartphone's resources without asking for permission could result in legal consequences for developers;
    
    \item To \textbf{protect} end-users (18\%): similar to our developer participants, some end-user participants also saw the protection of resources and their data as one of the purposes of permissions. They believed permissions could protect their personal data from unwanted access and protect their privacy; and
    
    \item To \textbf{collect} end-users' data (13\%): some end-user participants associated permissions with privacy-unfriendly practices. They thought apps requested permissions to collect data, track end-users, or target end-users with ads.
\end{itemize}

\subsection{Privacy Ramifications of Permissions}
\label{sec:privacy-implications}
\subsubsection*{\interviewTitle}
Most developer participants agreed that the \textbf{calendar} and all \textbf{location-related} permissions, especially when apps were \textbf{running in the background}, could invade end-user privacy because they enabled tracking and learning end-users' location at a specific time. While our end-user participants did not explicitly mention background and foreground services as a concern, prior literature suggests that background activity can influence end-users' decision-making processes with regard to granting or denying certain permission requests~\cite{shen2021can}.

Many developer participants expressed privacy concerns about the \textbf{location} permission because end-users could be tracked. It could also enable apps to \textbf{access metadata} and learn end-users' behavior by knowing, for example, whether the end-user was online or offline, as well as deduce the end-user location based on their network or Wi-Fi connection.

Several developer participants were also concerned about the \textbf{camera} and \textbf{microphone} permissions because they believed they allowed developers to capture end-users' surroundings, listen to all end-users' conversations, and record end-users without notification or feedback. Some were also concerned about the \textbf{contacts} permission because it could enable developers to sell end-users' contacts to third parties and generate profit. Although some developer participants recognized the benefits of accessing end-users' biometrics for facilitating authentication, they expressed concerns that the phone camera could capture unneeded data about end-users and their surroundings. Most developer participants agreed that the \textbf{Internet} permission was necessary for almost all apps to function correctly. However, it could invade end-user privacy if data was sent to \textbf{third parties} without end-users' knowledge or consent. It could also allow malicious actors on the Internet to track end-users or access their smartphones by exploiting unknown vulnerabilities (e.g., by using picture metadata or MAC address to locate end-users~\cite{reardon2019ways}).

Some developer participants also mentioned that Android had deprecated the \textbf{storage read} and \textbf{storage write} permissions due to security concerns (e.g., apps could access different account passwords stored in the same shared storage) and privacy concerns (e.g., apps should not share storage space; they should have their own private one). Some developer participants were surprised to see that granting permissions could give them \textbf{access to sensitive end-user data} and, hence, wanted to see \textbf{nuanced permissions} for additional end-user privacy: \blockquote[P5]{Some permissions are extremely open. Like, suddenly gives you a lot of access, which through the years, Android has fixed them actually. Like, when you wanted to read an SMS, you had to get SMS permission which suddenly allowed you to read all the SMS data on the phone, which they no longer allowed you to include that permission. Or, we have the get accounts permission, which again, if you want to introduce---add an account for your application to the device accounts---you have to add that permission. But that permission, at the same time, allows you to read all the other accounts that are defined on the device.}

Many developer participants thought that \textbf{security} testing, \textbf{security} measures, and data \textbf{security} \textbf{could provide privacy}: \blockquote[P8]{Nowadays, all the apps use stored data they have on a cloud system or something, so if I, as a developer, don't do my job right handling the security or permissions to access the accounts may leak the end-users later.} Some developer participants explained methods to \textbf{protect end-user privacy}: testing for security vulnerabilities, using multi-factor authentication, minimizing data collection, and giving end-users the option to limit data collection. For data protection and privacy reasons, few developer participants processed (or wished they could process) their data locally by \textbf{avoiding sending data to servers}.

\subsubsection*{\surveyTitle}

\begin{table*}
\centering
\caption{End-User participants' perspectives on harms associated with permissions.}
\Description{End-User participants' perspectives on harms associated with permissions. One column shows possible harms, and one shows themes when end-user participants did not see harm in permissions, respectively.}
\label{tab:end-users-reasons-harms}
\begin{tabular}{lrlr}
\toprule
\multicolumn{2}{c}{\textbf{Possible Harms With Permissions}} & \multicolumn{2}{c}{\textbf{Permissions Are Harmless}} \\ \midrule
Access to protected resources & 62\% & End-users are given a choice & 11\% \\
\quad Access to sensitive data & 18\% & No or minimal harm & 8\% \\
\quad Security issues & 18\% & End-users' responsibility & 5\% \\
\quad Selling data to third parties & 12\% & Trusting the developer & 3\% \\
\quad Unsure about data practices of developers & 3\% & Trusting the law & 2\% \\
Depending on the app & 6\% & Privacy resignation & 2\% \\ \bottomrule
\end{tabular}%
\end{table*}

Like developer participants, end-user participants viewed the location, photos, and storage permissions as sensitive with privacy ramifications (a lot or above: 68\%, 62.8\%, and 61.8\%, respectively). While the calendar permission came up as a privacy-sensitive permission request in the developer interview study, our end-user participants were not very concerned about it (a little or less: 48.2\%, see Appendix~\ref{app:plots-survey}, Figure~\ref{fig:sensitivePermsLikerts} for details).

In a series of Likert questions (Appendix~\ref{app:plots-survey}, Figure~\ref{fig:permsPrivacyLikerts}), we found that end-user participants were overall \textbf{worried} about permission requests. Many associated permission requests with tracking and monitoring (somewhat agree or above: 80.6\%), would think twice if the app requested too many permissions (somewhat agree or above: 66\%), and thought that apps with permissions used their data for \textbf{unintended purposes} (somewhat agree or above: 77\%).

\paragraph{Possible Harms With Permissions}
68.6\% of end-user participants said they could see possible harm in granting permission requests (Table~\ref{tab:end-users-reasons-harms}):

\begin{itemize}
    \item Access to \textbf{protected} resources (62\%): some end-user participants (18\%) were worried that apps could access their sensitive data (e.g., photos, location, and contacts). Some were also apprehensive about the unintended use cases of the permissions that they granted, such as security issues (18\%), selling their data to third parties (12\%), and being unsure about the data practices of developers (3\%); and
    
    \item \textbf{Depending} on the app (6\%): similar to the trust element discussed in both our developer and end-user studies, some end-user participants believed that the associated harms with permissions correlated with the developer's and app's reputation, echoing findings from prior work that brand reputation impacts end-users' decisions with regard to permissions~\cite{shen2021can}.
\end{itemize}

\paragraph{Permissions Are Harmless}
Below, we lay out the reasons end-user participants did not see any harm in granting permissions (31.4\%, Table~\ref{tab:end-users-reasons-harms}):

\begin{itemize}
    \item End-users are given a \textbf{choice} (11\%): giving end-users the option to grant or deny permissions made some incorrectly believe that this choice left them with no harm. If they were to use certain functionality, there would be no reason to consider other ramifications of granting permissions;
    
    \item \textbf{No or minimal} harm (8\%): on the other hand, some end-user participants incorrectly believed that permissions were often innocuous and that nothing bad was related to them. Few also thought that if there were no physical harm in granting permission, there would be no other harm;
    
    \item End-users' \textbf{responsibility} (5\%): similar to what some developer participants thought that granting permissions was an end-user choice and responsibility, few end-user participants also believed that it was their responsibility to choose the required permissions. Therefore, they thought no harm could happen because they could choose and assert their control over the app;
    
    \item \textbf{Trusting the developer} (3\%): depending on the app, how much end-user participants trusted the developer, and whether an app was from a well-reputed source, few end-user participants saw no harm in permissions;
    
    \item \textbf{Trusting the law} (2\%): while trusting the developer was a reason to believe an app was harmless, few end-user participants believed laws should protect them against any possible harm; and
    
    \item \textbf{Privacy resignation} (2\%): few end-user participants gave up on their privacy and said that their privacy would be invaded regardless of their decision. Few also said they had nothing to hide, so there was nothing to worry about.
\end{itemize}

\subsection{Reasons for Adding, Updating, or Removing Permissions}
\label{sec:reasons-adding}
\subsubsection*{\interviewTitle}
The reasons developer participants added, updated, and removed permission requests were:

\paragraph{Features \& Requirements} All developer participants agreed that app \textbf{features and requirements} were the primary reason for adding permissions. Based on what the client requested or the app required, developers added permissions. For example, a developer participant added the location permission because their client needed to keep track of parcels. Some developer participants acknowledged that adding \textbf{too many permissions} without a specific functionality or purpose could be \textbf{user-unfriendly}. Therefore, they kept their permission requests in proportion to the app's features.

\paragraph{Libraries}
To function correctly, some \textbf{libraries} required permissions, which was the second primary reason for developer participants to include permissions. We discuss the reasons for including or excluding libraries by developer participants in \S\ref{sec:reasons-libraries}.

\paragraph{Common Practice}
Many developer participants added permissions because of their everyday practices. The main reason was \textbf{reusing} a component in several projects because the apps they developed were similar, or projects \textbf{shared features}. For example, having a map in different apps required the inclusion of the location permission, or most apps needed to connect to a server; hence, including the Internet permission was considered standard practice. We also found that some developer participants copied and pasted code from the Internet, echoing prior work's findings~\cite{felt2011demystified}, or reused their own written code; both of these practices may leave unused or unnecessary permissions in the app.

\paragraph{Operating Systems \& App Stores}
\label{sec:os-appstore} 
Many developer participants updated their permissions to \textbf{satisfy operating system and app store requirements}. They changed the permissions they used due to changes in operating system permissions. Similarly, some received rejections from the Google Play Store or the Apple App Store after submitting their apps for review, making them change the permissions they initially added. We also explicitly asked developer participants about any changes they had to make to their app permissions to comply with \textit{privacy laws}. There was a mix-up among some developer participants over privacy laws and Apple's or Google's terms of service and privacy policies. However, all developer participants noted that they did not change their apps and permissions to specifically comply with privacy laws, such as the General Data Protection Regulation (GDPR)~\cite{gdpr2018eu}, CCPA, or COPPA. (However, investigating how developers comply with privacy laws was not the focus of our study. Therefore, we did not ask follow-up questions about this topic, see \S\ref{sec:limitations}.)

\paragraph{App Development Life Cycle}
Similar to what prior work has found~\cite{scoccia2019permissions, felt2011demystified}, some of our developer participants updated their permissions because of the \textbf{app development life cycle}. Some developer participants, for example, updated permissions due to requirement changes or app revisions. However, only a few developer participants mentioned they changed their permissions during testing as opposed to what prior work has suggested~\cite{felt2011demystified}. The difference could be due to the different research methods employed; \citeauthor{felt2011demystified}~\cite{felt2011demystified} analyzed Android apps (i.e., artifact analysis). We instead used direct retrospective interviews. Our developer participants might have changed their permissions during the testing and production stages but did not recall doing so during the interviews.

\paragraph{Code or App Crashes}
Some developer participants added permissions because they \textbf{received errors or warnings} from the development tool. As the developers' first point of contact, development tools can play a significant role in how developers write code and what they need to add to their code. Designing useful and enhanced compiler error messages shows promise and can assist novice developers in programming~\cite{becker2019compiler}. Therefore, lessons learned from prior studies could be used as a stepping stone to nudging developers toward permissions' privacy ramifications.

\paragraph{Workarounds}
\label{sec:workarounds}
Instead of adding permissions, few developer participants found other ways to access data \textbf{without asking end-users} or requesting permissions. For accessing location, for example, a few developer participants used the IP address instead of the location permission: \blockquote[P2]{The user will not give permission to disclose their location to the application . . . we can use other options to specify the regions specify the location. So, we tried to get the location from the IP.} 

One developer participant used the photo gallery permission instead of the camera permission because the gallery permission was viewed as an easier-to-get request, in their opinion. In this case, end-users were asked by the app to take a photo, save it to the photo gallery, and then share it with the app. Another developer participant had permission requests in the code but did not ask for them in earlier app revisions, so no permissions were requested when the app was released. In a later revision, they showed the permission requests to end-users when the permissions were needed. This was done by switching a Boolean flag in the code to true when a specific permission request was needed. Another developer participant considered using the list of installed apps on end-users' devices for fingerprinting; accessing the list of apps did not require a permission request and could be used to identify a device uniquely. The use of different methods to circumvent permissions is not new. Some large companies use multiple data points (e.g., network data) that do not require explicit end-user permission to infer end-users' location, for example~\cite{ftc2022mobile}.

\subsubsection*{\surveyTitle}
\label{sec:end-users-reasons-granting}
Most end-user participants \textbf{had not changed or rarely changed} their smartphone permission settings (never changed: 34\%; rarely changed: 35\%, Appendix~\ref{app:plots-survey}, Figure~\ref{fig:howOftensLikerts}). Below, we discuss end-user participants' reasons for granting and (not) removing permissions (Table~\ref{tab:end-users-reasons-removing}):

\begin{table*}
\centering
\caption{End-user participants' reasons (not mutually exclusive) for granting, removing, and not removing permissions.}
\Description{End-user participants' reasons for granting, removing, and not removing permissions. The figure has six columns; each column shows end-user participants' reasons for granting permissions, removing permissions, and not removing permissions with their respective percentages (themes are not mutually exclusive).}
\label{tab:end-users-reasons-removing}
\resizebox{\linewidth}{!}{%
\begin{tabular}{lrlrlr}
\hline
\multicolumn{2}{c}{\textbf{Reasons for Granting Permissions}} & \multicolumn{2}{c}{\textbf{Reasons for Not Removing Permissions}} & \multicolumn{2}{c}{\textbf{Reasons for Removing Permissions}} \\ \hline
Functionality \& features & 79.6\% & No need & 27\% & No longer needed & 21\% \\
App wouldn't work if perms weren't granted & 43.7\% & Comfortable in granted perms & 10\% & Feeling worried or uncomfortable & 16\% \\
Trusting the developer & 28.2\% & Did not know how & 8\% & Loss of trust & 4\% \\
Wanting the screen to go away & 18.4\% & Trusting the app & 6\% & Accidental perms & 4\% \\
Nothing to hide & 15.5\% & Functionality & 3\% & Deleted app & 2\% \\
Nothing bad would happen & 7.1\% & Forgotten perms & 2\% & Battery drainage & 2\% \\
Using popular app & 6.8\% & Deleted app instead of removing perms & 2\% &  &  \\
Developer already had the information & 4.9\% & No harm & 2\% &  &  \\
Wouldn't be able to grant later & 2.6\% &  &  & \multicolumn{1}{l}{} &  \\ \hline
\end{tabular}%
}
\end{table*}

\paragraph{Reasons for Granting Permissions}
The question about the reasons for granting permissions was from \citeauthor{bonne2017exploring} (2017)~\cite{bonne2017exploring} in which they did a longitudinal 6-week study with 157 participants to understand end-users' decision-making processes with regard to permissions. In the following paragraphs, we include percentages from our end-users and \cite{bonne2017exploring} for comparison; the first number in the parenthesis is ours, and the second number is from \cite{bonne2017exploring}.

Like our developer participants, end-user participants mentioned that the primary reason for granting permissions was \textbf{functionality} and features (79.6\% vs. 68.2\%), echoing prior findings~\cite{cao2021large,wijesekera2015android}. Else, they thought apps would not work if certain permissions were not granted (43.7\% vs. 23.8\%). \textbf{Trust} was the other reason for granting permissions (28.2\% vs. 32.1\%)---similarly, our developer participants wanted to build trust with end-user participants.

Related to the end-user experience theme from our developer interview study (\S\ref{sec:ux}), some end-user participants granted permission requests because they \textbf{wanted the screen to go away} (18.4\% vs. 10.2\%). A takeaway from our developer study is that developer participants tried not to flood the end-user with permission requests. However, some end-user participants were \textbf{overwhelmed} by permissions and wanted them to disappear (Figure~\ref{fig:devAssumptionsLikerts} also echoes this finding). Besides, some end-user participants said they granted permissions because they had \textbf{nothing to hide} (15.5\% vs. 18\%), and few thought that \textbf{nothing bad} would happen (7.1\% vs. 14\%).

The app's popularity was not a significant decision-making factor (6.8\% vs. 9.3\%); few end-user participants thought that developers already had data about them (4.9\% vs. 13\%)---suggesting that they viewed permissions as a guard against their data, and few noted that they could grant the permission later if they wanted to (2.6\% vs. 1.4\%). Figure~\ref{fig:whyChooseMultiple} in Appendix~\ref{app:plots-survey} shows the summary of the reasons for granting permissions.

\paragraph{Reasons for Not Removing Permissions}
\label{sec:reason-not-removing}
57.9\% of end-user participants had not ever removed permissions (Table~\ref{tab:end-users-reasons-removing}):

\begin{itemize}
    \item \textbf{No need} (27\%): some end-user participants never felt the need to remove permissions or thought it was unnecessary;
    
    \item Comfortable in \& \textbf{confident} about granted permissions (10\%): some end-user participants were happy with what they had already granted to the apps and did not see a reason to remove permissions;
    
    \item \textbf{Do not know how} (8\%): some end-user participants had no idea how to remove permissions;
    
    \item \textbf{Trusting the app} (6\%): some end-user participants trusted the app and did not see a reason for removing an already given permission;
    
    \item \textbf{Functionality} (3\%): the app provided functionality with the permissions. Therefore, only a few end-user participants saw a need to remove permissions, or else the app might break;
    
    \item \textbf{Forgotten permissions} (2\%): few end-user participants had not thought about removing permissions or had forgotten that they had granted permissions to an app that might need removing;
    
    \item \textbf{Delete} app instead of removing permission (2\%): few end-user participants decided to delete an app that had unnecessary permissions entirely instead of removing the unneeded permissions, which seemed to be a practice of few participants when they saw too many permissions or felt uncomfortable with permissions; and
    
    \item \textbf{No harm} (2\%): in line with associated harms, few end-user participants saw no harm in permissions and had no reason to remove an already granted permission.
\end{itemize}

\paragraph{Reasons for Removing Permissions}
42.1\% of end-user participants said that they had removed an already given permission from an app (Table~\ref{tab:end-users-reasons-removing}):

\begin{itemize}
    \item \textbf{No longer needed} (21\%): the primary reason to remove permissions was that the feature was no longer needed, such as giving permission to use photos for a certain time and later removing that permission;
    
    \item Feeling worried or \textbf{uncomfortable} (16\%): several end-user participants stated discomfort with having permissions left granted (e.g., contacts, location, and microphone);
    
    \item \textbf{Loss of trust} (4\%): following up on the trust theme described above in the developer and end-user studies, few end-user participants removed permissions because they heard some news about an app's privacy-unfriendly practices that made them lose their trust and, hence, removed permissions;
    
    \item \textbf{Accidental} permissions (4\%): few end-user participants granted permissions accidentally during installation or without carefully reading the permission descriptions. After a while, they realized the mistake and removed the permissions;
    
    \item \textbf{Deleted} app (2\%): few end-user participants also completely deleted the app when they found that permissions were unnecessary instead of removing permissions; and
    
    \item \textbf{Battery} drainage (2\%): few end-user participants also explicitly mentioned that they removed permissions because of battery consumption, especially when the permissions were related to some background activity.
\end{itemize}

\subsection{Considering End-Users \& Their Experience}
\label{sec:ux}
\subsubsection*{\interviewTitle\textbf{: Permission Descriptions}}
Many developer participants made an effort to write understandable descriptions. They acknowledged the value of \textbf{writing clear, transparent, and convincing descriptions} to help end-users understand why permission was needed and why access to a specific resource could help the app function correctly. Some developer participants were also motivated by app stores' requirements to provide clear and informative descriptions rather than short descriptions. Some developer participants wrote the description text, some received \textbf{help} from their product or design team, and others \textbf{tweaked predefined text} from online resources, operating systems, or app stores. Few developer participants had to localize their permission descriptions to \textbf{suit their end-user group}. They had end-users rejecting or blindly accepting permission requests because end-users could not understand the text.

Current permission dialogs only contain text, and some developer participants had to show \textbf{an extra page} ahead of the built-in permission dialog presented by the operating system to explain in more detail why the permission was requested: \blockquote[P3]{The official dialogue has minimal space to explain the user. So the additional dialogue helps the user understand more about the permission thing, and also, we can customize the dialog.} Although Apple recommends presenting end-users with a ``pre-alert screen'' before asking for sensitive permissions like location~\cite{apple2022privacyguide}, the additional screen or dialog may put an extra burden on the end-user because of extra clicks and interfaces (which may result in negatively impacting the app's rating~\cite{peruma2018investigating}) as well as the developer (e.g., extra work to create an additional dialog).

In 2014, in the early days of permission descriptions, some developers needed help understanding the value of these descriptions, saw them as unnecessary, and did not fully adopt them~\cite{felt2011effectiveness}. None of our developer participants questioned the existence of these descriptions; many even had favorable perspectives. The push from operating systems and app stores might be the driving force for such a shift in less than ten years, which echos other findings from privacy-related posts on developer forums that operating systems drive the app privacy ecosystem~\cite{tahaei2020so, li2021how}.

\paragraph{\surveyTitle}
End-user participants had \textbf{mixed opinions} about reading and checking permission descriptions (a little or less: 39.8\%, a moderate amount: 30.4\%, and a lot or more: 29.8\%, see Appendix~\ref{app:plots-survey}, Figure~\ref{fig:howOftensLikerts}).

\subsubsection*{\interviewTitle\textbf{: Assumptions About End-Users}}
\label{sec:assumptions}
We observed the following assumptions that developer participants made about end-users: (1) end-users are \textbf{only sensitive} about specific permissions and grant consent to others, (2) end-users \textbf{are not aware} of or do not know much about permissions, (3) end-users \textbf{do not care} about permissions as long as the app provides the expected functionality, and (4) \textbf{end-users are the responsible party} for any consequences of accepting permissions. For example, a developer participant viewed the location permission as easy to get, whereas the email permission as difficult to get, which could even dissuade end-users from using an app: \blockquote[P13]{If you give permission for the location, the user can easily allow them, but you can use the permission for messages and emails permission then some users have not provided that and they can choose the alternative app for that because the SMS and mail permissions are very sensitive for the users.} 

While the location permission can be viewed as sensitive by Google~\cite{google2022permissions}, getting this permission from our developer participants' perspective might be easy. Developer participants might understand the market and end-users' expectations by working with end-users. Permissions like contacts and messages (which may cause losing contact data or sending messages that could cost money) can be perceived by developers as more sensitive than the location permission~\cite{bonne2017exploring, felt2012ive}, which is a more classic privacy-sensitive permission from a research viewpoint~\cite{felt2012ive}.

\paragraph{\surveyTitle}
Most end-user participants \textbf{disagreed} with the developer participants' assumption that they \textbf{did not care} about permission requests (somewhat disagree or less: 81.2\%) or that there was no need for permission requests (somewhat disagree or less: 79\%). They strongly believed that permissions were necessary and were here to stay. Most end-user participants were sensitive about certain permissions, as some developer participants assumed about end-users. End-user participants also felt more comfortable granting a permission request if the app came from a known source, such as a large or well-reputed company (we discussed reputation and trust in \S\ref{sec:end-users-reasons-granting}). Most end-user participants \textbf{agreed} with the developer participants' assumption that it was their \textbf{responsibility} to grant or not grant a permission request (somewhat agree or more: 87.4\%). Nevertheless, many thought the current permission descriptions were broad and needed more precise and transparent descriptions (Figure~\ref{fig:devAssumptionsLikerts}).

\subsubsection*{\interviewTitle\textbf{: Request Permissions When Needed}}
Many developer participants tried to request permissions \textbf{when needed} or when there was a specific use case or functionality related to a permission request. They also acknowledged that end-users should be notified of accessing or using resources on their devices by the app. This involved thinking about the end-user experience and \textbf{trying not to flood the end-user} with many permission requests, as it could diminish the end-user experience and result in losing end-users.

Some developer participants also mentioned the importance of \textbf{trust and reputation} developers (or apps) built over time with end-users. Asking for \textbf{too many permissions} might result in \textbf{eroding trust} and negatively affecting the \textbf{end-user-developer relationship}. Maintaining this relationship could benefit developers in the long run, as end-user retention is one of the key determinants of a successful app~\cite{clever2022marketer, bizapps2022retention, ruba2017obstacles}. While we did not explicitly ask about breaching end-user trust, one developer participant admitted that they had \textbf{misused contact data} they collected in a personally-developed app by contacting someone they were not supposed to contact (they asked the contacted person to stop harassing one of the developer participant's family members). Such access would not have been given without the end-user trusting the app or its developer not to misuse their contact data.

\begin{figure*}
  \centering
  \includegraphics[width=.9\linewidth]{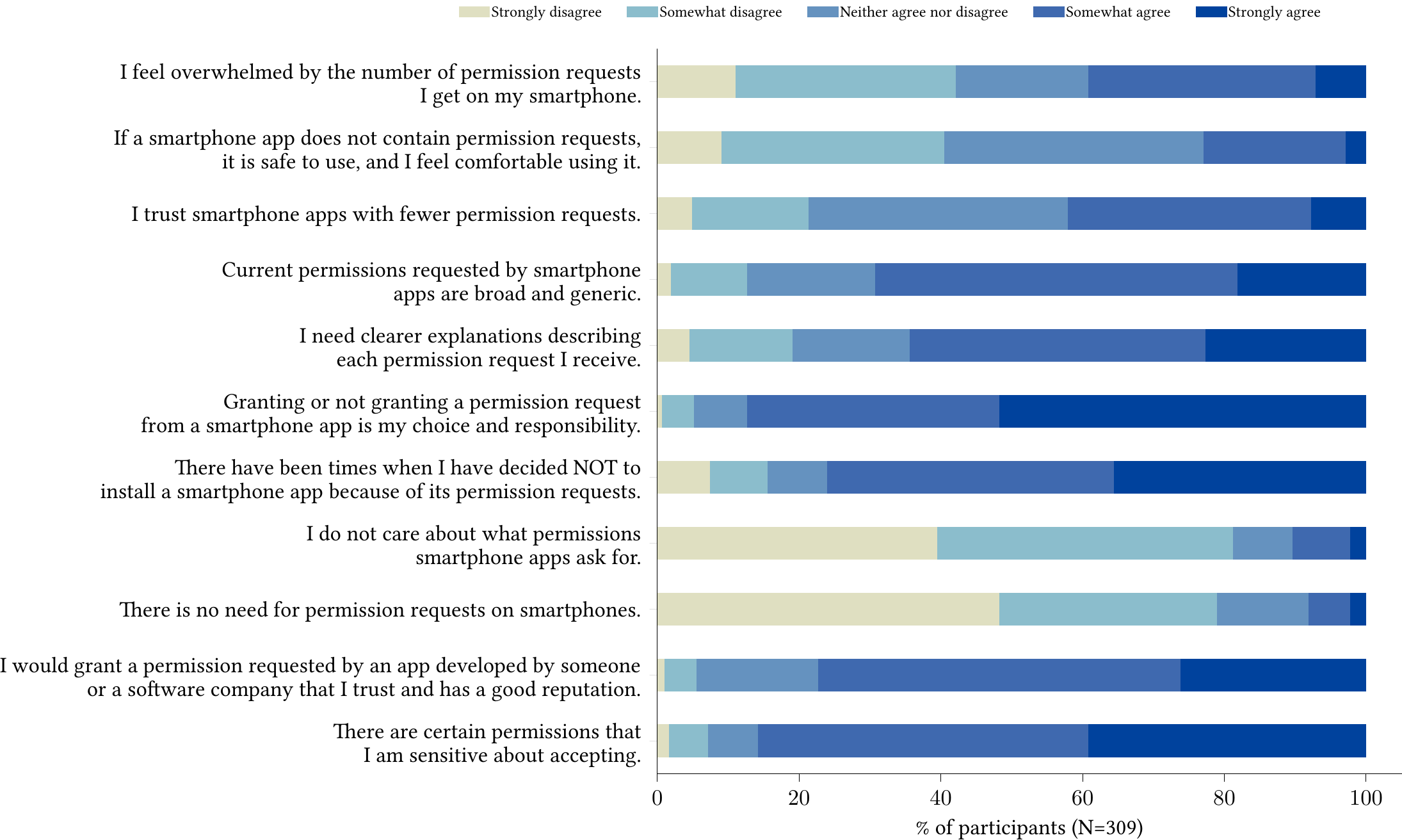}
  \caption{End-User participants' views compared with developer participants' views on and assumptions about end-users.}
  \Description{End-User participants' perspectives compared with developer participants' perspectives on and assumptions about end-users. Each bar shows end-users' answers to a Likert question.}
  \label{fig:devAssumptionsLikerts}
\end{figure*}

\paragraph{\surveyTitle}
Some end-user participants \textbf{trusted} apps with \textbf{fewer permissions} (somewhat agree or more: 42.1\%, somewhat disagree or less: 21.4\%, Figure~\ref{fig:devAssumptionsLikerts}). They also had mixed opinions about the association of permission requests with the app being perceived as safe or end-user participants feeling comfortable using it (somewhat agree or more: 23\%, somewhat disagree or less: 40.5\%). These mixed opinions may suggest that some end-user participants were aware of other ways of exploiting apps without requesting permissions (e.g., using an IP address instead of location data, as suggested by a few developer participants, \S\ref{sec:workarounds}).

\subsection{Developers' Challenges When Working With Permissions}
\label{sec:developer-challenges}
Most developer participants expressed satisfaction with the current permission integration mechanism. They mainly relied on Apple's and Google's documentation to learn more about permissions and clarify any confusion. They also consulted other resources such as Stack Overflow, GitHub, online tutorials and forums, YouTube, and search engines. Below, we discuss our developer participants' pain points in the process of permission integration:

\paragraph{Manual Changes \& Checks}
Many had to manually update permissions because of \textit{updates} to operating systems and libraries. Participants preferred that permissions were updated automatically without requiring manual changes. They also wanted libraries' permissions to get integrated automatically into the apps without manual involvement: \blockquote[P14]{That third party library is using it, but you have to ask for the permission by yourself.}

\paragraph{Late Feedback}
\label{sec:slow-feedback}
Developer participants often got notified of \textbf{additional or unexpected permissions} \emph{after} they developed their app, either from app stores or their testing or design teams. App rejection from app stores was the primary pain point, causing developer participants to wait for post-app submission feedback and then go back to the project to fix any newly flagged issues. In Android, some developer participants had to remove extra library permissions that they were unaware of because of receiving \textbf{warnings from Google Play}, suggesting that nudging developers toward reducing app permissions~\cite{peddinti2019reducing, google2020helping} seemed to be noticed by our developer participants and was viewed as a working mechanism: \blockquote[P5]{We offloaded the application on Google Play Store, and it showed some warnings about the permissions, and it was like, `I have never put these permissions inside the app. Where did this come from?' and then yeah, it was from a library. What I did was that I added some commands to just remove the unnecessary permissions, and the library worked.}

\paragraph{Confusing Scopes}
\label{sec:confusing-scopes}
When developer participants had to choose which permission to use for a feature, they were often \textbf{unsure} of what permission provided access to what resource. The mapping between the functionality required by an app and the needed permissions was a scoping issue that some developer participants had trouble understanding. It was even more challenging when \textbf{multiple} permissions were used for a \textbf{similar purpose}. For example, the location permission comes in various forms on both Android and iOS; developers can limit location data based on whether the app is running in the background or not, as well as the level of detail (e.g., coarse vs. fine)~\cite{apple2022locationPermissions, google2022locationPermissions}. Because of this confusion, some developer participants \textbf{included all} similar permissions to avoid missing a use case that required access to some needed data or resources and to prevent app crashes. P6, for example, wished for better documentation of permissions and their privacy ramifications: \blockquote[P6]{Documentation around what exactly each of these offer in terms of like sort of a reference functionality and perhaps including the privacy concerns around each of those.}

\paragraph{Poor Documentation}
Developer participants viewed the documentation provided by Android and iOS overall as \textbf{satisfactory}. However, some had issues understanding the documentation, \textbf{blindly trusted the sample code} given there, and wanted to see more code samples and guidelines explaining how to fix specific errors. A few developer participants stated that they fully trusted the documentation---setting a high standard for privacy from operating system providers---as these developers said they would copy-paste code samples from the documentation directly into their apps.

\paragraph{Not Informing Developers of Permission Changes by the End-User}
Few developer participants mentioned that they had trouble knowing \textbf{whether or not an end-user granted} permissions. In some cases, Apple's HealthKit, for example, does not inform developers of end-users' choices when (not) granting access to sensitive resources like health data to protect end-user privacy~\cite{apple2022protecting}.

\paragraph{Limited Support for Hybrid Development}
Few developer participants with hybrid app development experience (i.e., writing code once using a framework and exporting the app to several platforms) noted that they had to \textbf{keep track} of all the changes in Android and iOS, as well as understand the \textbf{different permission models} of the two operating systems.

\subsection{Developers' Reasons for (Not) Using Software Development Libraries}
\label{sec:reasons-libraries}

\paragraph{Functionality}
The primary reason for (not) using a library was its functionality, including features, ease of coding (\textbf{usability}), and adding control over code. Cross-platform libraries such as React Native and Flutter were the most mentioned libraries. In particular, some developer participants included libraries to \textbf{facilitate permission management}. Conversely, a library could have \textbf{too many features}, resulting in requesting \textbf{too many permissions}, which might result in the library getting removed from the project. In such scenarios, developer participants might decide to use an \textbf{alternative} library or write code from scratch.

\paragraph{Common Practice}
Some developer participants added a library because it was part of their \textbf{typical programming behavior}. Some basic libraries facilitated connection establishment, dependency management, permission management, image optimization, interface design, database management, and analytics.

\paragraph{Verified by Others}
When thinking about what libraries to include or exclude, some developer participants relied on what their \textbf{company or client approved}, \textbf{open-source} libraries hosted on GitHub, or libraries provided by operating system creators, \textbf{large technology companies} (e.g., Apple and Google), or a combination of these stakeholders.

Some developer participants viewed open-source libraries as reliable sources---even though Android repositories on GitHub could have permission-related issues, such as unused permissions~\cite{scoccia2019permissions}. A developer participant mentioned licensing issues with open-source projects that could complicate the use of such libraries.

\paragraph{Privacy Ramifications}
In some cases, developer participants were concerned about libraries' \textbf{complicated and murky practices} that they could not understand. Therefore, they decided to \textbf{reconsider} using such libraries. Third parties (e.g., ad networks and analytics services) could collect data from end-users without developers' knowledge, or the privacy settings of these libraries could be complicated and hard to find, putting an additional burden on developer participants~\cite{tahaei2022charting, alomar2022developers, reyes2018wont}: \blockquote[P19]{There is no way for us to know that a random SDK that we've pulled in they must be doing date changing or color monitoring---or it might be monitoring your app behind the scenes---actually we don't know. The whole thing is insanely complicated and gave me a really big headache.}

\paragraph{Rejection From App Stores}
A pain point for a few developer participants was getting an app rejected from app stores, as noted in \S\ref{sec:slow-feedback}. From a different perspective, this could make developer participants \textbf{review} their libraries and ensure that they did not collect \textbf{unnecessary} data or ask for unnecessary permissions: \blockquote[P1]{We submitted the app for Apple to review the app and then they said that the ad framework should not be there but if we remove the ad framework which we told them that if we remove the ad framework, Google Analytics will not work itself so ultimately, we had to remove Google analytics because there was no use of keeping the analytics without ad framework on for iOS.}

\paragraph{Considering End-Users \& Their Experience}
In some cases, developer participants decided to exclude a library because including it could have resulted in \textbf{poor end-user experience} (see \S\ref{sec:ux} for details about the end-user experience).

\section{Discussion and Future Work}

\subsection{Two Sides of the Same Coin: Developer \& End-User Perspectives on Permissions}
\label{sec:two-sides}
Both our developer and end-user participants acknowledged that privacy was at the heart of the app ecosystem, as well as that permissions were a key mechanism to protect end-users' privacy. Comparing the two stakeholders with the limitation that the data collection methods employed in this work are not the same gives us the following insights:

\subsubsection{Are Permission Requests Fully Grounded in App Functionality \& Features?\nopunct} 
Both stakeholders agreed that app functionality was the primary reason for including or granting permissions (\S\ref{sec:reasons-adding}). Similar to what prior studies have found, many of our end-user participants (79.6\%) tended to grant permissions necessary for the app to function, showing the importance of why permissions need to be contextualized and relate to functionality~\cite{micinski2017user, wijesekera2015android}. However, context is not static and, in practice, nor are permissions. End-users should be able to maintain permissions depending on context (they may deny a permission request in one context but let it run indefinitely in another until the need to revoke). This begs the question of whether permissions could help (empower) end-users to manage access to data and resources by a non-random app or whether they always need to grant specific permissions; otherwise, the app would not function.

On a similar note, we found that developer participants also made an effort to make permissions contextualized. However, developer participants might face challenges contextualizing all permissions because they could end up including more permissions than needed due to being confused about the scope of some permissions or fear of an app crash. Further, some developer participants mentioned that they needed to include certain third-party libraries in their apps, which requested more permissions than what apps needed, making contextualization difficult. One option would be for developers to write their own libraries, which undermines the concept of object-oriented software development and reusing software libraries.

Future work may want to employ the theory of Contextual Integrity~\cite{nissenbaum2004privacy} to explain, in a usable way, to end-users (or developers) the data flows that occur as a consequence of granting (or including) a permission request, giving more information to help make an informed decision. For example, end-users (or developers) may find denying (or excluding) a permission request by an app (or a third-party library) in certain contexts to be appropriate, even if they end up not being able to use the app (or library).

\subsubsection{Trust \& Reputation as Decision-Making Factors}
One recurring theme in both studies revolved around trust and reputation. Several developer participants recognized the value of building a relationship with end-users based on trust by requesting fewer permissions. Further, almost one-third of end-user participants trusted apps that asked for fewer or necessary permissions, echoing prior work findings that ``brand reputation'' is one of the factors influencing end-users' decision to grant or deny a permission request~\cite{shen2021can}.

\subsubsection{Who Is Responsible?\nopunct}
Based on our findings and prior work, we observed that developers, in practice, often shift the privacy ramifications of their development choices to end-users or platforms. Our results show that although several developer participants recognized the importance of not overwhelming end-users with unnecessary permission requests, they still thought end-users were ultimately responsible for deciding to (not) grant permissions requests. Further, most of our end-user participants believed this was their responsibility (\S\ref{sec:assumptions}). Prior work has also shown that developers often shift the ramifications of using software development platforms to platforms (e.g., ad networks~\cite{mhaidli2019cant, tahaei2021what}). Despite this belief, many software development platforms view developers as fully responsible for their code and apps (e.g., ad networks~\cite{mhaidli2019cant, tahaei2021what}, Apple App Store~\cite{apple2022userPrivacy}, and Google Play Store~\cite{google2022policyCenter}). Therefore, we believe this puts developers in a central place, as the \textit{mediators} between platforms and end-users, with their decisions directly impacting themselves and end-users.

There is a need to raise awareness in the developer community, perhaps through developer forums or academic educational platforms, to create a sense of responsibility and empathy toward end-users. Developers should not be the only responsible entity in the app ecosystem. Instead, responsibility should be shared across different entities (i.e., platforms, developers, and end-users) to reduce the load and burden on developers; exploring avenues for sharing this responsibility could be a future research direction.

\subsubsection{Permission Descriptions and Names Need a Fresh Look}
\label{sec:new-look}
Both developer participants and end-user participants agreed that transparent and clear permission descriptions and names (especially when asked at the right time~\cite{elbitar2021explanation}) would help end-users make informed decisions with regard to granting or denying a permission request (\S\ref{sec:ux}). About two-thirds of end-user participants, however, thought that current permission descriptions were broad and generic, expressing the need for more transparent and informative descriptions (Figure~\ref{fig:devAssumptionsLikerts}), confirming findings from \citeauthor{shen2021can}~\cite{shen2021can}. However, existing permission dialogs need more space to include informative explanations, forcing developers to create custom pages if they want to provide additional details. The extra pages can cause inconsistencies across apps and introduce an extra burden on developers. The usable security literature suggests that writing useful warning messages is a challenging task for developers~\cite{green2016developers}. Expecting developers to anticipate all possible contexts and use cases may be unrealistic. Therefore, future research may need to provide clear guidelines for developers on how to write consistent, contextual, transparent, and easy-to-digest permission descriptions (or experiment with visual metaphors) or improve the current permission dialogs of operating systems.

Similarly, some permissions have similar names but facilitate access to different types of data or resources on end-users' smartphones (e.g., coarse location, fine location), leading developers to get confused about the scope of such permissions. For example, several developer participants decided to include all permissions with similar scopes to ensure apps worked and did not crash (\S\ref{sec:confusing-scopes}). This could lead to dire privacy ramifications for end-users (who may also get confused about what a permission request does). We believe that although some permissions have similar scopes, they are still different, and clear and descriptive names that match the exact access type of a permission request are needed.

\subsubsection{Close the Gap Between End-Users' Beliefs, Comprehension \& Knowledge}
Most of our end-user participants believed that (1) they were aware of permission requests, (2) why permissions were requested, and (3) permission requests aligned with their expectations. However, we did not ask follow-up questions to test end-users' comprehension of different permission requests (see \S\ref{sec:why-perms} and Figure~\ref{fig:fiveAppsLikertsLikerts}). Besides our findings, a previous survey study has explored end-users' understanding of what different permission requests meant, finding that end-user participants struggled to correctly identify what a permission request did (only 6.1\% of participants answered all comprehension questions correctly)~\cite{shen2021can}. The difference between our findings and \citeauthor{shen2021can}'s findings may signal a gap between the ``beliefs'' and ``comprehension and knowledge'' components of the human-in-the-loop framework~\cite{cranor2008framework}. Future research may empirically investigate the consequences of this gap and suggest avenues to close this gap. As discussed in \S\ref{sec:new-look}, one possible improvement would be to provide clear permission names and descriptions.

\subsection{The Usual Suspects: Third-Party Data Practices Worry Developers \& End-Users}
A primary reason for including permissions by developer participants was to use third-party library features. Developers may feel a lack of power and control over third-party libraries, which could normalize the privacy-unfriendly practice of adding libraries requesting unneeded permissions. Future research is needed to understand the impact of normalizing privacy-unfriendly practices of developers on end-users. The literature on third-party privacy-unfriendly data practices is rich, and many studies have shown that third parties collect unwanted data from end-users, even sometimes without developers' knowledge~\cite{hallinan2021data, reardon2019ways, symeonidis2018collateral}. We believe that libraries need to break down their features into smaller pieces and enable developers to choose desired features. Only permissions needed by the chosen features can be included in apps instead of adding a large library with many unnecessary features.

We also observed in some instances that developer participants associated the privacy ramifications of libraries with permissions requested by libraries---thinking that if a library did not require permissions, it would be safe to use, and there would be no privacy concerns. This mindset is only partially accurate, as libraries can still collect data from end-users even when permissions are not requested (e.g., common libraries used for testing, analytics, and databases~\cite{hallinan2021data}). To address this complication, we suggest that third-party libraries transparently communicate their data collection practices to developers through easy-to-understand interfaces, especially since some of our end-user participants (\S\ref{sec:privacy-implications}) were concerned about their data getting harvested and sold to third parties. We also feel that apps should communicate to end-users the practices of these libraries. We speculate that apps that do so would have a competitive advantage over other apps that are not transparent, as privacy-conscious end-users may prefer to use apps that are advertised as privacy-friendly. However, this assumption requires further future research for validation.

\subsection{Forgotten Permissions: Possibly Rooted in Scope Confusions \& Lack of Awareness of Revoking Procedures}
We found that developer participants had several reasons for adding permissions, but they rarely had a reason to remove permissions (\S\ref{sec:reasons-adding}). Some said that they worried that removing permissions could cause an app crash or an error. Therefore, they would not want to try removing permissions from a working app---this is a common issue in apps~\cite{wang2022aper}. Some had trouble understanding the scope of some permissions leading to adding multiple permissions (e.g., location in use, location always) to avoid errors. Lessons learned from creating usable cryptography libraries for developers suggest that providing several implementation routes or using unfamiliar names for variables and functions may eventually lead to insecure actions by developers~\cite{green2016developers}. Similarly, as discussed in \S\ref{sec:two-sides}, we believe permission names and descriptions can benefit from clear naming and scopes to help developers decide on permission requests.

On the other hand, some end-user participants stated they had no idea how to remove permissions (57.9\%). While there has been research on how to present permission requests to end-users (e.g., \cite{wijesekera2018contextualizing, scoccia2018investigation}), we are unaware of specific studies that have explored ways to help end-users find and then remove unused or unnecessary permissions. Although some end-user participants were aware of permission settings and tried to change these settings (e.g., 28.68\% of Android participants, N=380~\cite{alsoubai2022permissions}), our findings suggest that more work is needed (\S\ref{sec:reason-not-removing}). Future research should consider informing end-users and developers of unused or unnecessary permissions periodically---one starting point could be using recommendation tools to periodically nudge and remind end-users and developers of permissions requested by apps, as inspired by \citeauthor{liu2016follow}~\cite{liu2016follow}.

A mechanism that notifies developers of unused and unnecessary permissions \textit{during} development (instead of post submission to app stores) or when releasing a revision could be a starting point to help developers. It could nudge developers toward removing unused permissions and remind them that end-users prefer apps with fewer permissions. For example, programming plugins for privacy checks (e.g., \cite{li2018coconut}) could implement such nudges for permissions. Relatedly, Android's recent addition to automatically check unused permissions for end-users and developers shows promise and highlights the importance of the problem~\cite{google2022revokePerm, google2022PermUpdates}. However, neither our developer nor end-user participants mentioned those checks, which could signal a lack of awareness. iOS, on the other hand, does not provide automatic revoking of unused permissions. Therefore, future research may examine the usability and effectiveness of these recent Android additions and procedures.

\section{Conclusion}
We interviewed 19 developers and surveyed 309 Android and iOS end-users to empirically study how these two stakeholders understand and interact with permissions. The mixed-methods design of our study allowed us to draw contrasting and diverse perspectives on permissions. Both developer and end-user participants associated permissions with app functionality and features. However, sometimes developer participants included multiple permissions due to confusion about the scope of certain permissions or third-party library requirements, which could worry both developers and end-users. We also discuss the implications of our findings and how to improve permissions for both developers and end-users.

\begin{acks}
We thank CHI23's anonymous reviewers, Pauline Anthonysamy, Tianshi Li, Ola Michalec, Marvin Ramokapane, Jose Such, William Seymour, and Xiao Zhan for their constructive feedback that helped improve the paper. The U.K.'s National Research Centre on Privacy, Harm Reduction, and Adversarial Influence Online (REPHRAIN) partly sponsored this research (EPSRC: EP/V011189/1).
\end{acks}

\bibliographystyle{ACM-Reference-Format}
\bibliography{main}

\appendix

\section{Screening Survey for the Interview Study With Developers}
\label{app:screening-survey}
[Answer options to close-ended questions were randomized where appropriate.]\\

You will have 30 to 60 seconds to answer each question on the next four pages. Each page states the time.

\begin{itemize}
    \item Which of these websites do you most frequently use as aid when programming?
    
    $-$ Wikipedia
    
    $-$ LinkedIn
    
    $-$ Stack Overflow
    
    $-$ Memory Alpha 
    
    $-$ I haven't used any of the websites above for programming
    
    $-$ I don't program
    
    \item Choose the answer that best fits the description of a compiler's function.
    
    $-$ Refactoring code
    
    $-$ Connecting to the network
    
    $-$ Aggregating user data
    
    $-$ Translating code into executable instructions 
    
    $-$ Collecting user data
    
    $-$ I don't know
    
    \item Choose the answer that best fits the definition of a recursive function.
    
    $-$ A function that runs for an infinite time
    
    $-$ A function that does not have a return value
    
    $-$ A function that can be called from other functions
    
    $-$ A function that calls itself 
    
    $-$ A function that does not require any inputs
    
    $-$ I don't know
    
    \item Which of these values would be the most fitting for a Boolean?
    
    $-$ Small
    
    $-$ Solid
    
    $-$ Quadratic
    
    $-$ Red 
    
    $-$ True
    
    $-$ I don't know\\

The remaining questions are NOT timed. \\

    \item Please answer the next two questions, given the following pseudocode algorithm.
    \begin{verbatim}   
    main {
        print(func("hello world"))
    }
    
    String func(String in ) {
        int x = len(in)
        String out = ""
        for (int i = x - 1; i >= 0; i--) {
            out.append(in [i])
        }
        return out
    }
    \end{verbatim}
    
    \item What is the parameter of the function?
    
    $-$ String out
    
    $-$ String in
    
    $-$ int i = x - 1; i >= 0; i-- 
    
    $-$ Outputting a String 
    
    $-$ int x = len(in)
    
    $-$ I don't know
    
    \item Please select the returned value of the pseudocode above.
    
    $-$ hello world
    
    $-$ hello world 10
    
    $-$ dlrow olleh
    
    $-$ world hello 
    
    $-$ HELLO WORLD
    
    $-$ hello world hello world hello world hello world
    
    $-$ I don't know

    \item Please select the statement that best describes your primary role at your current or most recent job. 
    
    $-$ Jobs NOT related to computer science, informatics, computer engineering, or related fields
    
    $-$ Designing products (e.g., UI designer, interaction designer)
    
    $-$ Developing software (e.g., programmer, developer, web developer, software engineer)
    
    $-$ Testing software (e.g., tester, quality analyst, automation engineer)
    
    $-$ Managing software development (e.g., project manager, IT manager, scrum master)
    
    $-$ Privacy and/or security engineering (e.g., security engineer, privacy engineer, penetration tester, ethical hacker, cryptographer)
    
    $-$ Other (please specify)
    
    \item Are you a student?
    
    $-$ Yes, I'm a student in computer science or related fields
    
    $-$ Yes, I'm a student but NOT in computer science or related fields
    
    $-$ No, I'm NOT a student

    \item How many years of experience do you have in software development? [Numbers only]
    
    \item How many years of experience do you have in Android programming? [Numbers only]
    
    \item How many years of experience do you have in iOS programming? [Numbers only]
    
    \item How many years old are you? [Numbers only]
    
    \item In which country do you currently reside? [List of countries]

    \item If you can't find your country in the above question options, please enter it here. [Open-ended question]
    
    \item What is your gender?

    $-$ Male 
    
    $-$ Female 
    
    $-$ Non-binary 
    
    $-$ Prefer not to say 
    
    $-$ Prefer to self describe
    
\end{itemize}

\section{Demographics for the Interview Study With Developers}
\label{app:demographics-developer}
Table~\ref{tab:demographics} shows a summary of developer demographics.

\begin{table*}
\centering
\caption{A summary of developer participant demographics.}
\Description{A summary of developer participant demographics. Each column shows a demographic variable and the rows represent each of the developer participants.}
\label{tab:demographics}
\begin{tabular}{lcccclll}
\toprule
\textbf{PID} & \textbf{\begin{tabular}[c]{@{}c@{}}Yrs of expr \\ in software dev\end{tabular}} & \textbf{\begin{tabular}[c]{@{}c@{}}Yrs of expr \\ in Android dev\end{tabular}} & \textbf{\begin{tabular}[c]{@{}c@{}}Yrs of expr \\ in iOS dev\end{tabular}} & \textbf{Age} & \textbf{\begin{tabular}[c]{@{}l@{}}Continent \\ of residence\end{tabular}} & \textbf{Gender} & \textbf{\begin{tabular}[c]{@{}l@{}}Recruitment \\ platform\end{tabular}} \\ \midrule
P1 & 10 & 3 & 7 & 35--44 & Asia & Male & Freelancer \\
P2 & 5 & 5 & 4 & 25--34 & Asia & Male & Freelancer \\
P3 & 5 & 5 & 4 & 25--34 & Asia & Male & Freelancer \\
P4 & 5 & 4 & 0 & 25--34 & Europe & Male & LinkedIn \\
P5 & 22 & 13 & 0 & 35--44 & Asia & Prefer not to say & LinkedIn \\
P6 & 6 & 3 & 2 & 25--34 & North America & Male & Prolific \\
P7 & 22 & 1 & 10 & 35--44 & North America & Male & Prolific \\
P8 & 5 & 0 & 4 & 18--24 & Asia & Male & Freelancer \\
P9 & 5 & 5 & 1 & 25--34 & Asia & Male & Freelancer \\
P10 & 7 & 0 & 5 & 25--34 & Asia & Female & Freelancer \\
P11 & 8 & 8 & 1 & 25--34 & Asia & Male & Upwork \\
P12 & 3 & 3 & 0 & 25--34 & Africa & Female & Upwork \\
P13 & 5 & 5 & 5 & 25--34 & Asia & Male & Upwork \\
P14 & 5 & 2 & 5 & 25--34 & Asia & Male & Upwork \\
P15 & 5 & 3 & 1 & 25--34 & Africa & Male & Upwork \\
P16 & 12 & 6 & 6 & 35--44 & South America & Male & Upwork \\
P17 & 7 & 5 & 1 & 25--34 & Europe & Prefer not to say & Upwork \\
P18 & 6 & 4 & 1 & 25--34 & Africa & Male & Upwork \\
P19 & 8 & 8 & 8 & 25--34 & Europe & Female & LinkedIn \\ \bottomrule
\end{tabular}%
\end{table*}

\section{Interview Guide for the Interview Study With Developers}
\label{app:interview-guide}
\begin{itemize}
\item
  Can you tell me about your job? What do you do?
\item
  Can you tell me about the apps that you have developed?

  \begin{itemize}
  \item
    What kind of apps do you develop?
  \item
    What are the age groups of your users?
  \item
    What kind of data do you collect in your apps?
  \end{itemize}
\item
  Have you ever needed to share data between different apps? [If
  yes]

  \begin{itemize}
  \item
    Why, and how?
  \item
    Has it been between the apps you developed or between other apps and
    your apps?
  \end{itemize}
\item
  What types of permissions do you often include in your apps?
\item
  Is there a set of permissions you often include, or do you pick
  permissions per app?
\item
  How do you decide on which permissions to include and which
  permissions not to include?
\item
  How do you think permissions work?
\item
  Do you ever update the permissions of your apps? [If yes]

  \begin{itemize}
  \item
    Why?
  \item
    How frequently?
  \item
    Can you think of an example?
  \end{itemize}
\item
  Have you ever removed permissions? If so, why?
\item
  What is the most confusing thing you have experienced when working with permissions?

  \begin{itemize}
  \item
    What sources did you consult to sort out the confusion?
  \end{itemize}
\item
  How do you think your choices of permissions impact your users'
  choices of apps?

  \begin{itemize}
  \item
    What do you think the reasons are behind users accepting or
    rejecting permissions?
  \item
    How do you think as a developer you can better guide your users to accept permissions?
  \end{itemize}

  \begin{itemize}
  \item
    How do you decide on what to include in the descriptions?
  \item
    How does adding descriptions impact your work?
  \end{itemize}
\item
  Do you include third-party libraries or SDKs in your apps? [such as
  ads, logging tools, and analytics] [If yes]

  \begin{itemize}
  \item
    How often?
  \item
    Can you name a few that you often include in your apps? For what
    functionality do you use the SDKs?
  \item
    How do you decide on permissions of these libraries?
  \item
    How do you know what permissions the library needs?
  \item
    Have you ever decided not to include a library because of its
    permissions?
  \end{itemize}
\item
  Now, I am going to send you a list of permissions in the chat. [We showed participants the 15 most commonly-used permissions in Android, iOS, or both, taken from \citeauthor{kollnig2022are}~\cite{kollnig2022are}, based on participants' platform expertise.] I would like
  you to have a look at them and tell me in your opinion how each
  permission behaves, or what it does, when you include the permission in your
  app.

  \begin{itemize}
  \item
    Please go back to the list and tell me which ones do you think may
    have a privacy consequence for your users? Why, and how?
  \end{itemize}
\item
  What data do you consider private or sensitive for your users?
\item
  Have you ever heard about any privacy regulations?

  \begin{itemize}
  \item
    Have you considered making any of your apps compliant with any
    privacy regulations? If yes, why and how?
  \item
    Have you made any changes to your permissions because of a privacy
    regulation?
  \end{itemize}
\item
  Can you think about any potential harms of permissions to your users?

  \begin{itemize}
  \item
    Can you think of any privacy consequences of permissions for your users?
  \end{itemize}
\item
  How would you want to see permissions managed in mobile apps to make
  it easier for you to manage and integrate permissions?
\item
  If you were to redesign the permission models, how would you do it?

  \begin{itemize}
  \item
    What information and support should be included in mobile operating
    systems to help you better understand the language, options, and
    interfaces of permissions?
  \end{itemize}
\item
  Would you like to share any other experiences related to our
  conversation today, specifically about permission models in smartphone operating systems?
\end{itemize}

\section{Codebook for the Interview Study With Developers}
\label{app:codebook}
Table~\ref{tab:codebook} shows the codebook and the number of unique participants mentioning each theme.

\begin{table*}
\centering
\caption{Codebook for the interview study with developers. Occurrences show the unique number of participants in each theme.}
\Description{Codebook for the interview study with developers. Occurrences show the unique number of participants mentioning each theme. The table has four columns, two for themes and two for occurrences of themes.}
\label{tab:codebook}
\resizebox{\linewidth}{!}{%
\begin{tabular}{lrllr}
\toprule
\textbf{Theme} & \textbf{Occurrences} &  & \textbf{Theme} & \textbf{Occurrences} \\ \midrule
\textit{Reasons for adding, updating, or removing permissions} & 19 &  & \textit{Reasons for (not) using software development libraries} & 19 \\
    \quad Features \& requirements & 19 &  & \quad Reasons for exclusion & 14 \\
    \quad Libraries & 17 &  &   \quad \quad Permission management & 14 \\  
    \quad Common practice & 16 &  & \quad \quad Extra or unnecessary features & 6 \\
    \quad Operating systems \& app stores & 15 &  & \quad \quad \quad Adds or loses control over code & 5 \\
    \quad App development life cycle & 7 &  & \quad \quad  Privacy implications & 6 \\
    \quad Code or app crashes & 5 &  & \quad \quad App rejection from app store & 5 \\
    \quad Workarounds & 4 &  & \quad \quad End-users and UX & 5 \\
     &  &  & \quad Reasons for inclusion & 19 \\
    \textit{Developers' challenges when working with permissions} & 19 &  & \quad \quad Functionality & 19 \\
    \quad Manual changes \& checks & 15 &  & \quad \quad \quad Adds control over code & 5 \\
    \quad Late feedback & 13 &  & \quad \quad Common practice & 14 \\
    \quad Confusing scopes & 12 &  & \quad \quad  Usable & 13 \\
    \quad Poor documentation & 9 &  &  \quad \quad Verified by others & 12 \\ 
    \quad Not informing developers of permission changes by the end-user & 5 &  & \quad \quad \quad Company, client-approved & 6 \\
    \quad Limited support for hybrid development & 2 &  & \quad \quad \quad Open-source & 6 \\
     &  &  & \quad \quad \quad Operating systems or large tech companies & 4 \\
    \textit{Considering End-Users \& their experiences} & 19 &  &   \quad \quad Permission management & 6 \\
    \quad Permission descriptions & 18 &  &  &  \\
    \quad \quad Localization of permission descriptions & 3 &  & \textit{Privacy conceptualization} & 19 \\
    \quad \quad Who writes permission descriptions & 15 &  & \quad Security implies privacy & 14 \\
    \quad \quad \quad Developer writes text & 12 &  & \quad \quad Privacy measures & 8 \\
    \quad \quad \quad Predefined text for perms & 5 &  & \quad Nuanced permissions & 7 \\
    \quad \quad \quad Someone else provides the text & 6 &  & \quad Processing data locally & 3 \\
    \quad Assumptions about end-users & 14 &  &  &  \\
    \quad \quad End-Users are sensitive about certain permissions & 12 &  & \textit{Why do permissions exist?} & 19 \\
    \quad \quad End-Users don't know (aren't aware) & 7 &  & \quad Access control & 19 \\
    \quad \quad End-Users don't care & 5 &  & \quad \quad User consent to accessing a resource & 18 \\
    \quad Shifting responsibility to end-users & 12 &  & \quad \quad Terms like ``turning on or off a switch'' & 7 \\
    \quad Request permissions when needed & 15 &  & \quad \quad As a key to a door & 3 \\
    \quad \quad End-Users should be notified & 11 &  & \quad \quad As a bridge & 2 \\
    \quad \quad Trust \& reputation & 9 &  &  &  \\ \bottomrule
\end{tabular}%
}
\end{table*}

\section{Questionnaire for the Survey Study With End-Users}
\label{app:end-users-survey}
[Answer options to close-ended questions were randomized where appropriate.]

\begin{itemize}
    \item On average, how many hours do you spend on your primary smartphone on a daily basis? Please estimate your daily average usage in hours. [Slider 0--24]
    
    \item For how many years have you been using a smartphone? (Numbers only)
    
    \item Please name five smartphone apps that you have frequently used in the past year? Use commas to separate the items. For example, app1, app2, app3, app4, app5.

    \item In this survey, we use the term ``permissions'' or ``permission dialogs'' to refer to the following example dialogs in smartphone operating systems like Android and iOS. [Example dialogs from Android and iOS documentation were included]
    
    \item In one sentence, why do smartphone apps request permissions? There is no right or wrong answer, please use your own words, as we are not looking for a correct or a perfect answer.

    \item In one sentence, how do you explain what permissions requested by apps do in smartphones? There is no right or wrong answer, please use your own words, as we are not looking for a correct or a perfect answer.
    
    \item Please select the top five permission requests that you have frequently seen in the past year on your smartphone? 
    $-$ Photos or Gallery 
    $-$ Location 
    $-$ Camera 
    $-$ Internet 
    $-$ Contacts 
    $-$ Microphone 
    $-$ Sensors or Motion 
    $-$ Calendars 
    $-$ Bluetooth 
    $-$ Biometrics (e.g., Touch ID or Face ID) 
    $-$ Storage or files 
    $-$ WiFi or Network state 

    \item If a smartphone app uses permissions, it means that the app . . .
    
    $-$ Can see the content of all the files on the device you are using 
    
    $-$ Is not a risk to infect your device with a computer virus 
    
    $-$ Will automatically prompt you to update your web browser software if it is out of date 
    
    $-$ Can access certain resources on your device 
    
    $-$ I don't know 

    \item Why do you choose to grant permissions requested by smartphone apps? (Select all that apply)
    
    $-$ I want to use a specific feature that requires permission
    
    $-$ I think the app will not work otherwise 
    
    $-$ I trust the app developer 
    
    $-$ Because the app is popular 
    
    $-$ I will not be able to grant permissions later 
    
    $-$ I have nothing to hide 
    
    $-$ I want the permission screen to go away 
    
    $-$ Nothing bad will happen 
    
    $-$ The app developer already has the requested 
    information about me 
    
    $-$ I don't know 
    
    \item To what extent do you agree or disagree with the following statements? [Scale: Strongly disagree, Somewhat disagree, Neither agree nor disagree, Somewhat agree, Strongly agree]
    
    $-$ There are certain permissions that I am sensitive about accepting. 
    
    $-$ I do not care about what permissions smartphone apps ask for. 
    
    $-$ There have been times when I have decided NOT to install a smartphone app because of its permission requests. 
    
    $-$ Granting or not granting a permission request from a smartphone app is my choice and responsibility. 
    
    $-$ I need clearer explanations describing each permission request I receive. 
    
    $-$ Current permissions requested by smartphone apps are broad and generic. 
    
    $-$ I trust smartphone apps with fewer permission requests. 
    
    $-$ If a smartphone app does not contain permission requests, it is safe to use, and I feel comfortable using it. 
    
    $-$ I feel overwhelmed by the number of permission requests I get on my smartphone. 
    
    $-$ I would grant permission requested by an app developed by someone or a software company that I trust and has a good reputation. 
    
    $-$ There is no need for permission requests on smartphones. 

    \item Thinking about the five smartphone apps that you have frequently used in the past year, to what extent do you agree or disagree with the following statements? [Scale: Strongly disagree, Somewhat disagree, Neither agree nor disagree, Somewhat agree, Strongly agree]
    
    $-$ I am aware of the permissions used in the smartphone apps that I frequently use. 
    \balance
    $-$ Permissions requested by the smartphone apps that I frequently use are aligned with my expectations. 
    
    $-$ I understand why certain permissions are requested by smartphone apps that I frequently use. 

    \item Have you ever removed a permission that you have granted?
    
    $-$ Yes, I have removed a permission that I granted before. Please explain why and how.
    
    $-$ No, I have not ever removed a permission that I granted before. Please explain why.
    
    \item Do you think permissions requested by a smartphone app you use could harm you?
    
    $-$ Yes, I can see harm in permission requests. Please explain why
    
    $-$ No, I do not see any harm in permission requests. Please explain why.
    
    \item In your opinion, to what extent does granting the following permissions on your device have privacy implications for you? [Scale: None at all, A little, A moderate amount, A lot, A great deal]
    
    $-$ Photos or Gallery 
    $-$ Location 
    $-$ Camera 
    $-$ Internet 
    $-$ Contacts 
    $-$ Microphone 
    $-$ Sensors or Motion 
    $-$ Calendars 
    $-$ Bluetooth 
    $-$ Biometrics (e.g., Touch ID or Face ID) 
    $-$ Storage or files 
    $-$ WiFi or Network state 

    \item Please rate the following statements. [Scale: None at all, A little, A moderate amount, A lot, A great deal]
    
    $-$ How often do you check the permissions requested by an app before installing it on your smartphone? 
    
    $-$ How often do you read the description and explanation of permission dialogs? 
    
    $-$ How often do you contact the developer of a smartphone app to ask for more information and clarifications about the app's permissions? 
    
    $-$ How often do you change the settings of your smartphone to manage the permissions  of a smartphone app? For example, to remove an already given permission or to grant an already removed permission. 
    
    \item To what extent do you agree or disagree with the following statements? [Scale: Strongly disagree, Somewhat disagree, Neither agree nor disagree, Somewhat agree, Strongly agree]
    
    $-$ After installing an app on my smartphone, it would bother me if the app started asking me to grant permissions. 
    
    $-$ If I installed an app on my smartphone, I would think twice before granting any of the app permissions. 
    
    $-$ If I accepted app permissions, my activities on my smartphone would be monitored or tracked at least part of the time. 
    
    $-$ If I accepted app permissions, I would be concerned that the app would know more information about me. 
    
    $-$ If I accepted app permissions, I would be concerned that the app would monitor my activities on my smartphone. 
    
    $-$ If I accepted app permissions, others would know about me more than I would be comfortable with. 
    
    $-$ If I accepted app permissions, information about me that I consider to be private or sensitive would be more easily accessible to others than I would want. 
    
    $-$ If I accepted app permissions, information about me would be out there. Also, if that information were  used, my privacy would be invaded. 
    
    $-$ If I accepted app permissions, I would be concerned that the app might use my personal information for other purposes without notifying me or getting my authorization. 
    
    $-$ If I accepted app permissions, I would be concerned that the app might use my information for other purposes. 
    
    $-$ If I accepted app permissions, I would be concerned that the app might share my personal information with other entities without getting my authorization. 
    
\end{itemize}

\section{Additional Plots for the Survey Study With End-Users}
\label{app:plots-survey}

\begin{figure*}
  \centering
  \includegraphics[width=.6\linewidth]{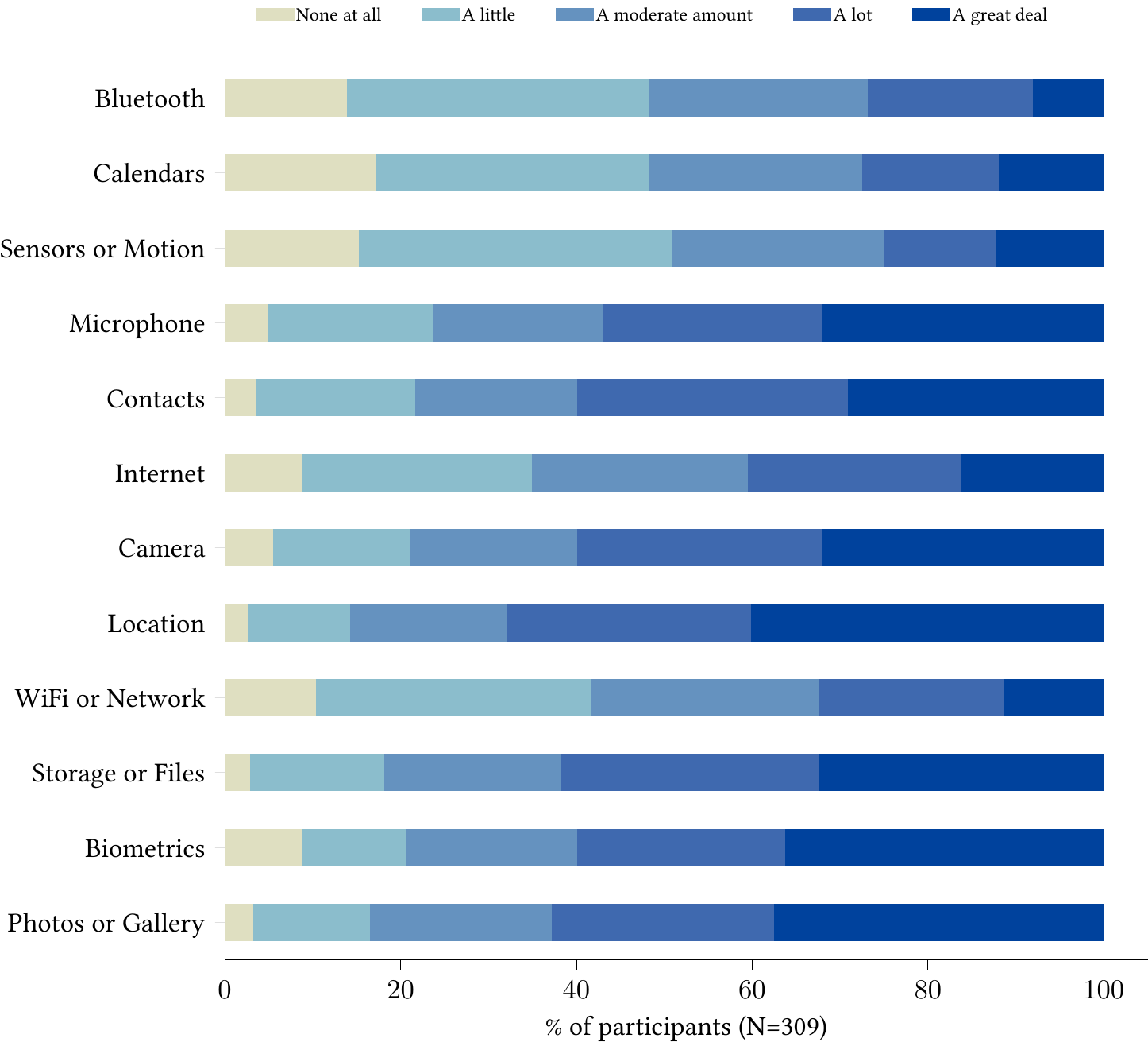}
  \caption{End-User participants' answers to ``In your opinion, to what extent does granting the following permissions on your device have privacy implications for you?''}
  \Description{End-User participants' answers to ``In your opinion, to what extent does granting the following permissions on your device have privacy implications for you?'' Each bar shows end-users answers to a Likert question.}
  \label{fig:sensitivePermsLikerts}
\end{figure*}

\begin{figure*}
  \centering
  \includegraphics[width=\linewidth]{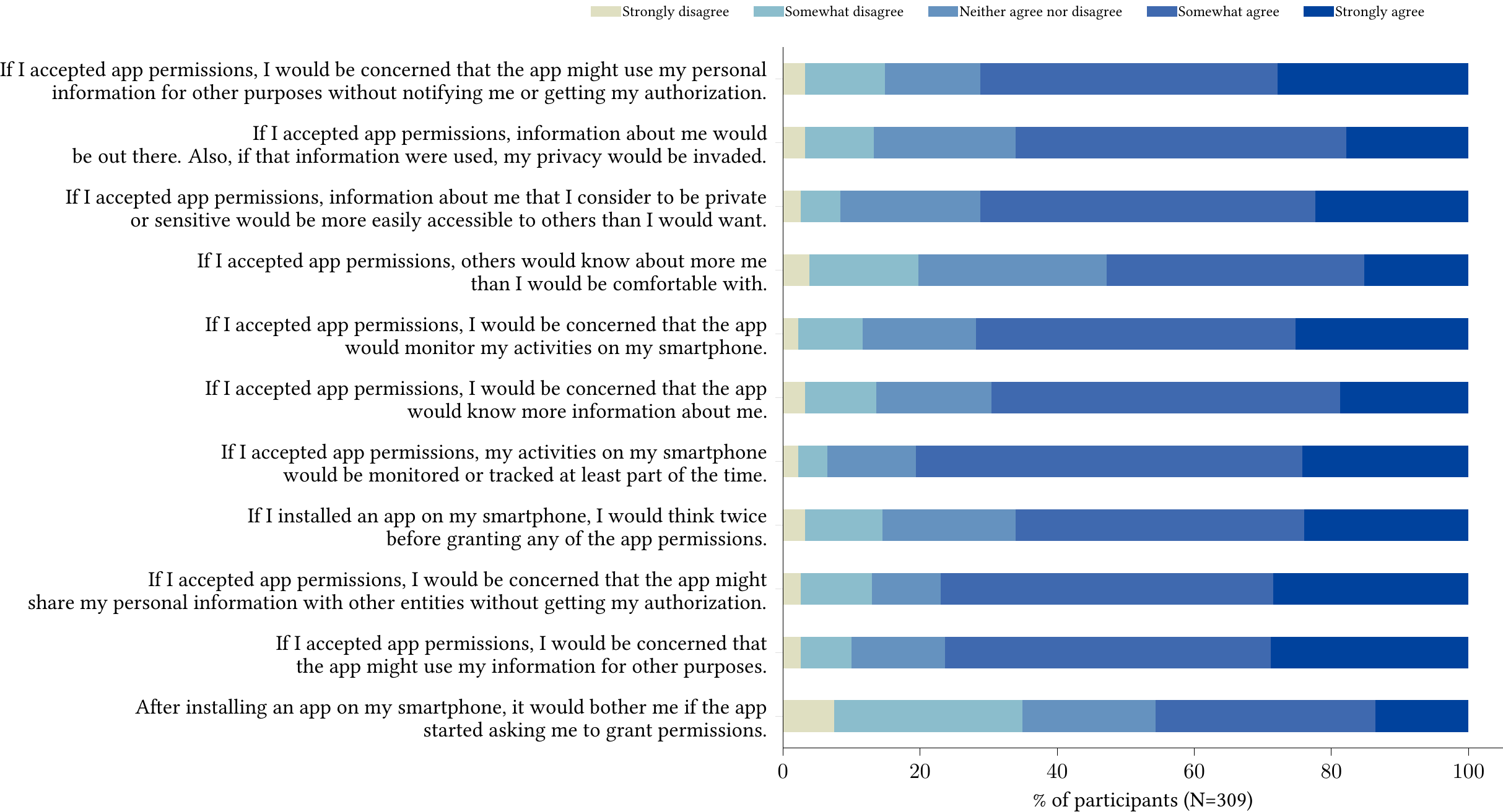}
  \caption{End-User participants' answers to questions about privacy ramifications of permissions.}
  \Description{End-User participants' answers to questions about privacy ramifications of permissions. Each bar shows end-users answers to a Likert question.}
  \label{fig:permsPrivacyLikerts}
\end{figure*}

\begin{figure*}
  \centering
  \includegraphics[width=\linewidth]{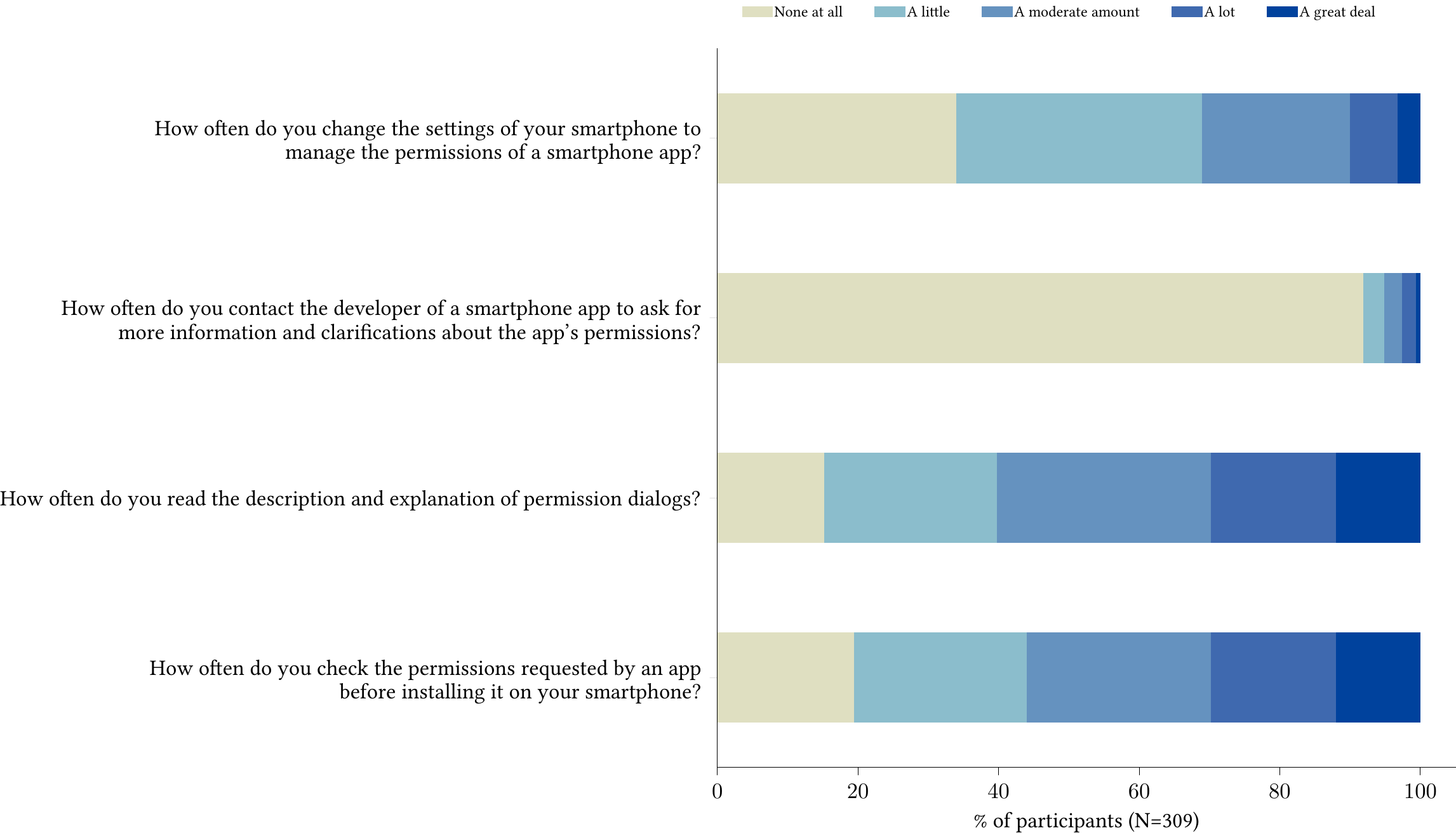}
  \caption{End-User participants' answers to questions rooted in the developer study.}
  \Description{End-User participants' answers to questions rooted in the developer study. Each bar shows end-users answers to a Likert question.}
  \label{fig:howOftensLikerts}
\end{figure*}

\begin{figure*}
  \centering
  \includegraphics[width=.6\linewidth]{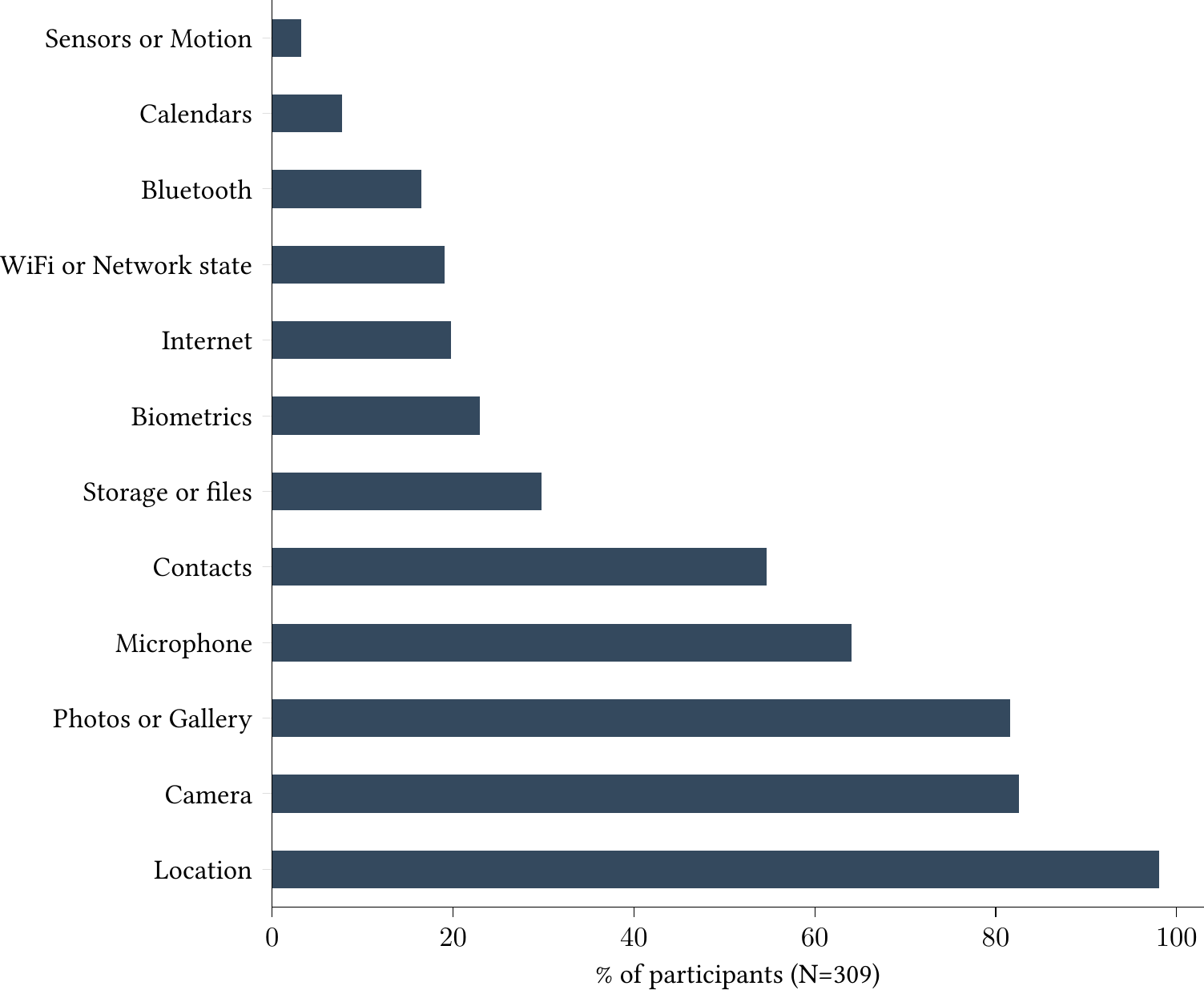}
  \caption{End-User participants' answers to ``Please select the top five permission requests that you have frequently seen in the past year on your smartphone?''}
  \Description{End-User participants' answers to ``Please select the top five permission requests that you have frequently seen in the past year on your smartphone?'' Each bar shows end-users answers to a Likert question.}
  \label{fig:fivePerms}
\end{figure*}

\begin{figure*}
  \centering
  \includegraphics[width=.8\linewidth]{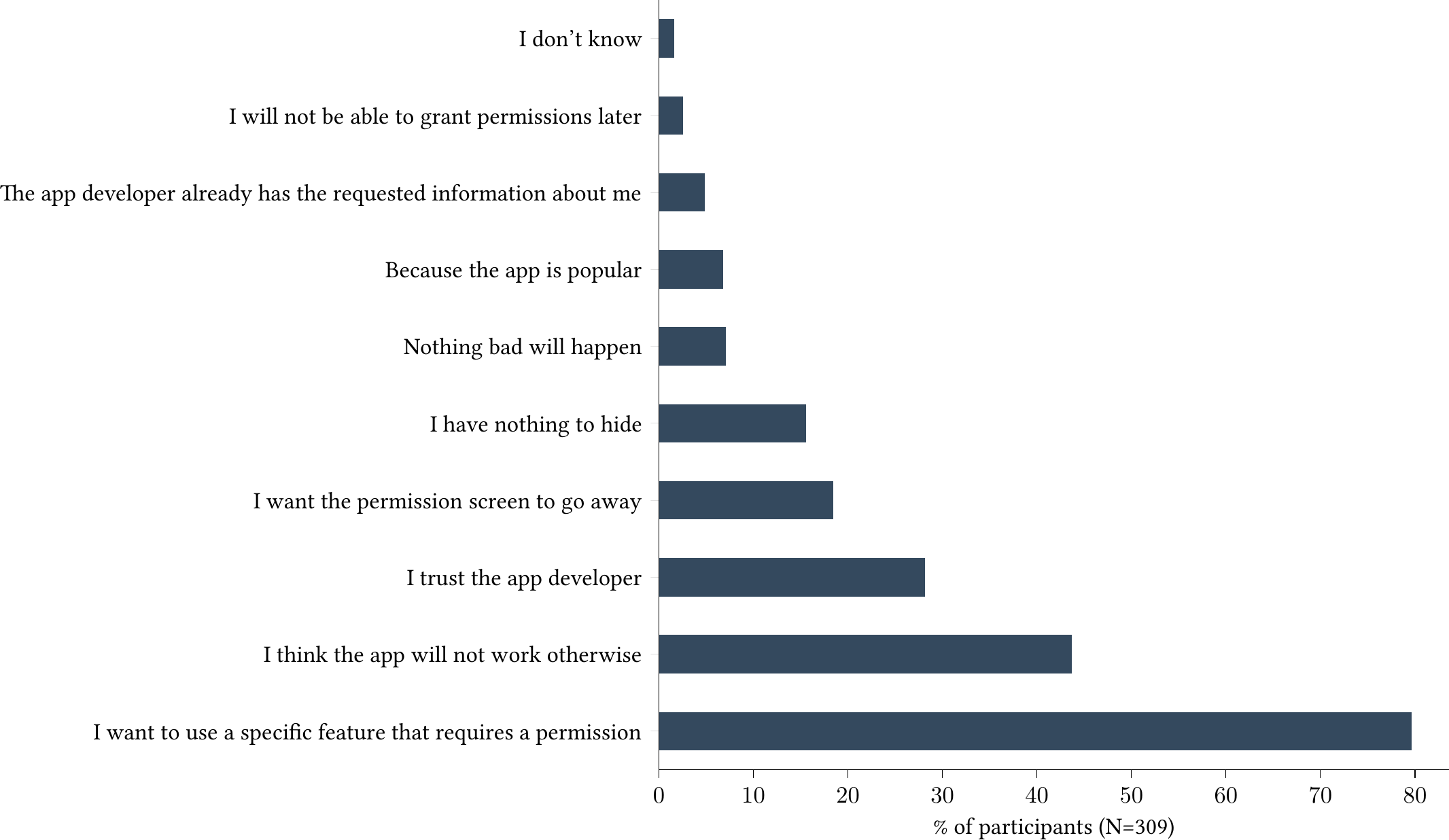}
  \caption{End-User participants' answers to ``Why do you choose to grant permissions requested by smartphone apps? (Select all that apply).''}
  \Description{End-User participants' answers to ``Why do you choose to grant permissions requested by smartphone apps? (Select all that apply).''}
  \label{fig:whyChooseMultiple}
\end{figure*}

\end{document}